\documentclass[journal]{vgtc}                
\ifpdf
  \pdfoutput=1\relax                   
  \usepackage{graphicx}                
  \DeclareGraphicsExtensions{.pdf,.png,.jpg,.jpeg} 
\else
  \usepackage{graphicx}                
  \DeclareGraphicsExtensions{.eps}     
\fi%

\graphicspath{{figures/}{pictures/}{images/}{./}} 

\usepackage{microtype}  
\usepackage{caption}
\usepackage{subcaption}

\PassOptionsToPackage{warn}{textcomp}  
\usepackage{textcomp}                  
\usepackage{mathptmx}                  
\usepackage{times}                     
\usepackage{cite}                      
\usepackage{tabu}                      
\usepackage{booktabs}
\usepackage{multirow}

\onlineid{1649}

\vgtccategory{Research}
\vgtcpapertype{Analytics \& Decisions (Area 6)}

\title{Intuitively Assessing ML Model Reliability through \\ Example-Based Explanations and Editing Model Inputs}

\author{Harini Suresh\\\scriptsize{hsuresh@mit.edu} %
\and Kathleen M. Lewis\\\scriptsize{kmlewis@mit.edu} %
\and John V. Guttag\\\scriptsize{guttag@mit.edu}
\and Arvind Satyanarayan\\\scriptsize{arvindsatya@mit.edu}
}
\affiliation{\scriptsize MIT CSAIL}

\shortauthortitle{Intuitively Assessing ML Model Reliability}
\shortauthortitle{Suresh \MakeLowercase{\textit{et al.}}: Intuitively Assessing ML Model Reliability}

\abstract{
Interpretability methods aim to help users build trust in and understand the capabilities of machine learning models. However, existing approaches often rely on abstract, complex visualizations that poorly map to the task at hand or require non-trivial ML expertise to interpret. Here, we present two visual analytics modules that facilitate an intuitive assessment of model reliability. To help users better characterize and reason about a model’s uncertainty, we visualize raw and aggregate information about a given input’s nearest neighbors. Using an interactive editor, users can manipulate this input in semantically-meaningful ways, determine the effect on the output, and compare against their prior expectations. We evaluate our interface using an electrocardiogram beat classification case study. Compared to a baseline feature importance interface, we find that 14 physicians are better able to align the model's uncertainty with domain-relevant factors and build intuition about its capabilities and limitations.
} 

\keywords{Interpretability, Machine learning, Visualization, K-nearest neighbors, Example-based explanations}

\CCScatlist{ 
 \CCScat{K.6.1}{Management of Computing and Information Systems}%
{Project and People Management}{Life Cycle};
 \CCScat{K.7.m}{The Computing Profession}{Miscellaneous}{Ethics}
}

\teaser{
  \centering
  \includegraphics[width=\textwidth]{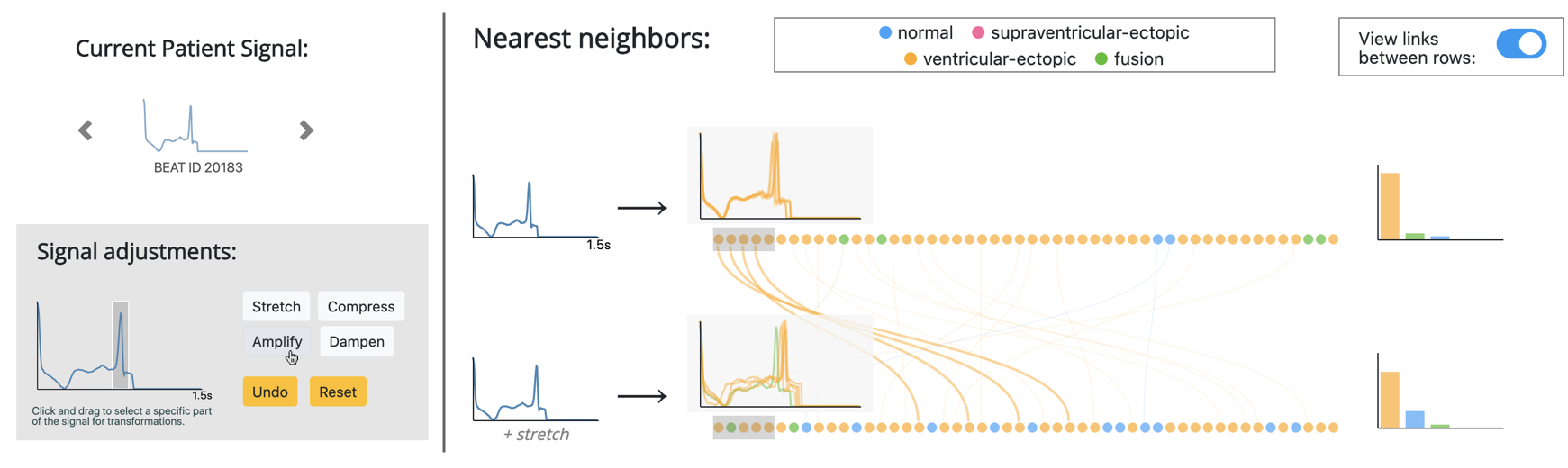}
  \caption{An example of the proposed interface for an electrocardiogram (ECG) case study. The output of the machine learning model consists of raw and aggregate information about the input's nearest neighbors. With the editor in the bottom left, the user can apply meaningful manipulations to the input and see how the output changes.}
  \label{fig:teaser}
}

\renewcommand{\manuscriptnotetxt}{}

\begin{document}


\firstsection{Introduction}

\maketitle

Machine learning (ML) systems are being developed and used for a broad range of tasks, from predicting medical diagnoses~\cite{jiang2017artificial} to informing hiring decisions~\cite{ajunwa2016hiring}. 
Many are intended to be part of a larger sociotechnical process involving human decision-makers.  
In these cases, in-domain accuracy is not enough to guarantee good outcomes\,---\,the people using a particular system must also understand the model's reliability (i.e., when its predictions should be trusted, in general and on a case-by-case basis) and modulate their trust appropriately~\cite{selbst,jacovi2021formalizing}.  
\textit{Model interpretability}, which is broadly intended to give insight into how a particular ML model works, can play an important role here.  

Many existing approaches to model interpretability, however, require a non-trivial amount of ML expertise to understand, and thus are often only used in practice by ML developers~\cite{bhatt_explainable_2020}.  
While tools for developers are certainly needed, the people who will actually deal with model predictions during decision-making are often a distinctly different set of users. Even methods that are intended to be simpler and more understandable to such users, such as reporting feature weights or displaying more information about the model and dataset, have not improved decision-making in experimental studies \cite{poursabzi-sangdeh_manipulating_2019, lai_human_2019, bussone_role_2015, sureshMisplaced,jesus2021can}.

Here, we introduce two visual analytics modules to facilitate more intuitive assessment of model reliability. 
First, we use k-nearest neighbors (KNN) to ground the model's output in examples familiar to the user~\cite{renkl_toward_2014}. 
Alongside the overall distribution of neighbors, a unit visualization depicts individual example, encoding their class and similarity to the original input according to the model. An interactive overlaid display provides a more raw visualization of the examples for more detailed comparison.
Second, we introduce an interactive editor for probing the model. Users can apply transformations corresponding to semantically-meaningful perturbations of the data, and see how the model's output changes in response. 
Using these modules together, users can iteratively build their intuition about the model's strengths and limitations. 
By interactively examining individual neighbors, they can investigate questions like whether variation amongst the neighboring examples is expected for the domain, or if it indicates unreliability; whether the commonalities amongst neighbors align with domain knowledge; or whether these neighbors reveal limitations or biases in the data. 
Similarly, by interactively modifying the model's input, users can pose and test hypotheses about the model's reasoning, checking that its behavior aligns with domain expectations\,---\,for example, ensuring that the model is not overly sensitive to small input modifications that should be class-preserving.

We evaluate the effectiveness of our interface modules through a medical case study of classifying electrocardiogram (ECG) heartbeats with different types of irregularities. 
This case study allows us to perform an application-grounded evaluation~\cite{doshi-velez_towards_2017} with representative real-world decision-makers who have prior knowledge and investment in the domain.  
We conducted think-aloud studies with 14 physicians, observing the way they interacted with our interface as well as a feature importance baseline. 
When working with the baseline, participants often rationalized incorrect predictions\,---\,for example, back-tracking on their initial assessment and seeking out things in the input that justified the model's incorrect prediction.
In contrast, the KNN visualizations help participants grasp prediction reliability\,---\,for example, by being able to determine whether variations between neighbors was the result of natural ambiguities in ECG data, or whether it reflected the model not learning the right representations for the task. 
Moreover, by exploring neighbors from different classes, participants were consistently able to relate the model’s uncertainty to clinically-relevant concepts to guide decision-making\,---\,for example, pulling out higher-level pathologies that differed amongst neighbors from different classes to understand why the model would be split between those classes. 
Finally, participants used the input editor to iteratively form hypotheses about the model’s reasoning and test them, using the results to investigate how the model worked and whether its reasoning was clinically sensible.  

Our proposed visual analytics modules contribute to the growing work on designing human-centered interfaces for ML systems that highlight both model strengths \textit{and} weaknesses, and that encourage critical engagement with the system. We highlight several important design goals to this end, including grounding visualizations in examples familiar to the user, enabling comparison across examples, and allowing interactive probing of the model. Importantly, we align visual components and modes of interaction with users' existing conceptual models of the domain, and show that this facilitates more intuitive understanding of the model and its reliability. Our work suggests several promising directions for future research aiming to improve human-ML interaction, from better conveying data limitations upfront to balancing user input with automated methods when probing the model.   

\section{Related Work}

\subsection{Interpretability Methods for Human Understanding}
ML interpretability aims to provide information that helps people understand how a model works, either on a global or case-by-case level \cite{gilpin_explaining_2018}.  Such efforts can serve a number of different goals, such as aiding in decision-making, helping debug or improve a system, or building confidence in the model \cite{hong_human_2020}.  A major area of focus has been on developing methodologies for computing and presenting such explanations \cite{carvalho_machine_2019}.  

Some methods try to visualize the internals of a particular model to reason about how it is operating \cite{michelini_multigrid_2019, carter_exploring_2019,zeiler2014visualizing}. This can be useful for theoretical ML understanding and model development, but may be too abstract and complicated to help people without knowledge of such models and how they work.  Others try to produce explanations more grounded in the features of the data, such as a ranking of features important for the prediction or a decision-tree approximating the model’s logic \cite{du_techniques_2019, ribeiro_why_2016, lundberg_unified_2017}. However, a growing body of work that has tried to empirically measure the efficacy of many of these methods has shown that they often do not actually affect or improve human decision-making \cite{poursabzi-sangdeh_manipulating_2019, lai_human_2019, adhikari_leafage_2019, bussone_role_2015,jesus2021can}, and in practice are primarily used for internal model debugging \cite{bhatt_explainable_2020}.

To understand the discord between proposed interpretability methods and their suitability for real-world users, we can draw from well-established theories in cognitive psychology that describe how people think about problems and organize information using different “cognitive chunks” \cite{miller1956magical}.  For example, a physician might think about diagnostic decisions in terms of concepts that are higher-level than individual features, or relate features to each other in more complex ways than independently ranking them by importance.  This idea manifests in theories of HCI stating that effective and engaging interfaces should allow users to view and interact with them in a way that feels \textit{direct}\,---\,i.e., the visualizations and interactive mechanisms available to users should align with their cognitive chunks. Specifically, Hutchins et al. \cite{hutchins1985direct} describe “the gulf of execution,” arising from a gap between the available mechanisms of an interface and the user’s thoughts and goals, and “the gulf of evaluation,” arising from a gap between the visual display of an interface and the user’s conceptual model of the domain.  Our aim is to narrow both of these gaps.

To this end, example-based (also referred to as instance-based) interpretability methods, which produce explanations in terms of other input examples, are of particular interest.   Research in cognitive psychology and education supports the idea that people often use past cases to reason about new ones when solving problems \cite{aamodt_case-based_1994} and that utilizing examples can help people understand complex concepts, build intuition, and form better mental models \cite{renkl_example-based_2009,renkl_toward_2014}.   

Different types of example-based explanations for ML models have been proposed. Many of these are computed \textit{post hoc}, i.e., they are generated after a prediction is made to try and explain that prediction. For example, counterfactual examples \cite{wachter_counterfactual_2018, goyal_counterfactual_2019} use gradient-based methods to generate the closest example(s) to the input that are predicted to be a different class (defining appropriate measures of “closeness” is an open question).  Influence functions \cite{koh_understanding_2017}  try to trace a model’s predictions back to the data it was trained on, identifying the examples that were most influential to the prediction.  Normative explanations \cite{cai_effects_2019} present users with a set of training examples from the predicted class. There are limitations of these approaches as well; for example, technical constraints make quickly generating influential examples quite difficult in practice \cite{basu_influence_2020, bhatt_explainable_2020}, and hidden assumptions about actionability in counterfactual explanations can be misleading \cite{barocas_hidden_2020}.  

Others compute example-based explanations by modifying the inference process of a trained model to produce predictions based directly on similar training examples. For example, both Caruana et al. \cite{caruana1999case} and Shin and Park \cite{shin1999memory} use a trained neural network model to improve a KNN classifier, either through using the model to create a weighted similarity function or through computing similarity in the embedding space of the model, respectively.  The class label making up the majority of nearest neighbors can be interpreted as the prediction, and the nearest neighbor examples used as an explanation.  Of particular relevance to our case study, Caruana et al. \cite{caruana1999case} is motivated by the potential benefits of example-based explanations in clinical settings: “\textit{because medical training and practice emphasizes case evaluation, most medical practitioners are adept at understanding explanations provided in the form of cases}.”  Recently, Papernot and McDaniel \cite{papernot_deep_2018} extended this methodology to compute neighbors using embeddings from multiple layers of a neural network, demonstrating additional uses for improving the model’s robustness and confidence estimates.  

In our proposed interface, we compute neighbors using the method of \cite{caruana1999case}; this could be easily extended to calculate neighbors in a weighted input space as in \cite{shin1999memory}, or to use embeddings from multiple layers of the neural network as in \cite{papernot_deep_2018}. Prior work has focused on developing optimal ways for the trained neural network to inform the KNN algorithm, implying that the nearest neighbors would then serve as an explanation.  Here, we focus on a relatively unexplored part of this claim, investigating how the resultant output should be presented to the user in an interactive interface to narrow the gulfs of execution and evaluation.  We explore a specific case study to more clearly define the ways in which this type of example-based explanation can improve trust and understanding for real users.     
  
\subsection{Interactivity and Visualization for Interpretability}
For interpretability to be useful in practice, effectively communicating information to the user is a critical step. In a literature review of interpretability systems and techniques, Nunes and Jannach \cite{nunes2017review} found that the vast majority of papers presented explanations in a natural-language-based format (e.g., a list of feature weights). Other types of visualizations include simple charts (e.g., bar plots indicating feature importances) \cite{ribeiro_why_2016} or highlighting/denoting sections of the input (e.g., displaying important pixels of an image in a different color or opacity) \cite{sturmfels_visualizing_2020, lai_human_2019}. With respect to example-based explanations, the visualizations used are often a table of features if the data is tabular \cite{wachter_counterfactual_2018,mothilal_explaining_2020,wexler2019if} or a list of images if the data is image-based \cite{koh_understanding_2017, kim_examples_2016,cai_effects_2019}.  Here, we explore visual encodings that convey more information and allow for more interaction than listing examples.

Other work specifically focuses on visualizations of latent embeddings within a neural network model.  Many of these utilize 2 or 3D plots to visualize distance between different examples in the embedding space \cite{liu2019latent,boggust2019embedding,heimerl2018interactive}. Liu et al. \cite{liu2019latent} additionally visualize examples along 1D vectors corresponding to user-defined concepts, and Boggust et al. \cite{boggust2019embedding} provide the ability to compare embeddings of two different models by viewing and interacting with the two plots side-by-side.  Particularly relevant to our work, some of the visualizations of text embeddings proposed in \cite{heimerl2018interactive} aim to display a given word’s nearest neighbors in an embedding space. They plot the nearest neighbors as points along a 1D axis that encodes distance, and provide the ability to compare the nearest neighbors across different embeddings.  

With respect to interactivity in the interface, prior work has primarily studied using human feedback to modify or filter the information that is shown \cite{kim_interactive_2015, kulesza_principles_2015,sokol2020one,cai2019human}.  Here, our goal is instead to provide users with a way to probe the model and test hypotheses about its behavior.  The tool described in \cite{wexler2019if} similarly allows modifying the input to observe how a model’s output changes, though in their case, it is intended primarily for users familiar with ML.  

Like these prior works, our interface aims to facilitate understanding by allowing users to visualize and interact with examples from the data. However, while they are primarily intended for general exploration of what a model has learnt, or for uncovering underlying structure in data, the goal of our interface is to help users assess the reliability of predictions on a case-by-case basis.

\section{Visual Analytics for Intuitive Model Assessment}

We introduce two visual analytics modules for intuitively assessing the reliability of ML models.  
In Sec. \ref{sec:design_goals}, we outline the goals that guide our designs. 
The proposed modules utilize general ideas that can be customized to different domains, and we illustrate them with a concrete instantiation of the ECG beat classification task introduced in Sec. \ref{sec:case_study}.
We then describe the visual components of each module: a display of the model’s output in terms of an aggregate and an individual-level view of nearest neighbors (Sec. \ref{sec:knn}), and an editor with which users can interactively modify model input and observe how the output changes in response (Sec. \ref{sec:editor}). Finally, in Sec. \ref{sec:use_cases} we walk through specific ways that users can interact with the visual analytics modules to more intuitively assess the model and its predictions. 

\subsection{Design Goals}\label{sec:design_goals}

To facilitate intuitive assessment of model behavior, our overarching goal is to narrow the \textit{gulf of evaluation} and \textit{gulf of execution} for the users of our visual analysis interfaces~\cite{hutchins1985direct}.  We identify several sub-goals to this end, which motivate the design of our interfaces:

\begin{itemize}
    \item [\textbf{G1.}] \textbf{Ground visualizations in examples.} To narrow the gulf of evaluation, the visual components of our interface should facilitate reasoning that aligns with users' existing conceptual models. We draw from research suggesting that reasoning through prior examples can aid in problem-solving \cite{aamodt_case-based_1994}, understanding, and mental model-building over time \cite{renkl_example-based_2009,renkl_toward_2014}. Particularly for users who are more familiar with the application domain than the mechanisms of ML models, using examples is likely to facilitate more intuitive reasoning than approaches based on model components or individual features (consider reasoning about anatomical structures in an x-ray versus individual pixels). Therefore, we aim to use real examples as the building blocks of our visualization. 
    
    \item [\textbf{G2.}] \textbf{Facilitate comparisons across examples.} To further facilitate interaction more aligned with users' existing modes of thinking, we are motivated by literature suggesting that \textit{contrastive} reasoning (i.e., reasoning based on what makes a particular case different than similar cases) is a particularly important way that people understand and explain things \cite{miller2019explanation,lipton1990contrastive}. Building on this, we aim to make it straightforward for users to compare specific examples in terms of meaningful high-level concepts in the data, enabling them to build understanding with contrastive reasoning.
    
    \item [\textbf{G3.}] \textbf{Visualize distributions over predicted classes.} Often, the output of ML-based systems is just a single predicted class, which may convey a false sense of certainty and prompt over-reliance, as some studies have found \cite{gaube2021ai,lee2004trust}. 
    On the other hand, conveying model uncertainty can help users align model behavior with their understanding of inherent challenges or ambiguities in the task \cite{cai2019hello,tonekaboni2019clinicians}. Indeed, research on human trust suggests that in addition to conveying assurances of certainty, acknowledging when systems are \textit{un}certain is also an important factor in building effective trust \cite{jacovi2021formalizing}.
    Providing a probability score along with the prediction is one way to convey uncertainty, though understanding how to interpret abstract probability values is itself challenging for people.  Instead, we aim to visualize the output from the model as a distribution over classes at multiple levels of granularity.  For example, visualizing an overall probability distribution alongside the specific examples belonging to each class may help users better grasp the sources of the model’s (un)certainty and reconcile it with their own understanding of the task.  
    
    \item [\textbf{G4.}] \textbf{Enable interactive probing of the model in terms of domain-relevant concepts.} 
    Prior work interviewing ML stakeholders has found that one way to build trust is to provide users with ways to confirm that the model is using sensible logic that aligns with their expectations \cite{bhatt_explainable_2020, hong_human_2020,liao2020questioning,tonekaboni2019clinicians,cai2019hello}. 
    To facilitate this process, we are motivated by the call to design for ``contenstibility,'' i.e., to make questioning and probing the model an integrated part of the system, rather than ``out-of-band activities'' \cite{mulligan2019shaping,hirsch2017designing}. 
    Interactive capabilities for exploring and querying the model can encourage this kind of engagement\,---\,prompting a back-and-forth process where users develop hypotheses and test them, confirming that the model’s behavior aligns with their domain knowledge or uncovering unexpected issues. 
    To minimize the gulf of execution, it is also important that users can form such queries in terms of domain-relevant and semantically-meaningful concepts. 
\end{itemize}

\subsection{ECG Beat Classification Case Study }\label{sec:case_study}

Although our visual analysis interfaces are general-purpose and can be adapted for different domains, we will use a specific case study of classifying electrocardiogram (ECG) beats as a case study to more concretely instantiate and evaluate our ideas. This task allows us to perform an application-grounded evaluation of our system using a realistic task\footnote{Using an overly-simplified or proxy task can be more straightforward, but can also yield less reliable results as it is not something the participants are familiar with or have prior conceptions about doing \cite{bucinca2020proxy}.} that people (i.e., physicians) are familiar with \cite{doshi-velez_towards_2017}.
ECG beat classification, in particular, is an area where machine learning has been widely applied and yielded good performance \cite{sannino2018deep, zubair2016automated, kachuee_ecg_2018}.  It is also generally applicable in the medical domain since most physicians are familiar with reading ECG beats.

The specific task we implement is classifying a single ECG heartbeat into one of four categories: normal, supraventricular ectopic, ventricular ectopic, or fusion.  The latter three classes are different types of arrhythmias, or heart rhythm problems. 
We use a preprocessed version of the MIT-BIH Arrhythmia Dataset \cite{moody2001impact} available on Kaggle \cite{kaggle_dataset}.  Each sample in the dataset is an individual heartbeat sampled at a frequency of 125 Hz, and padded to a maximum length of 1.5 seconds.  The available dataset contains a fifth class, ``unknown,'' which we exclude here.
We replicate the convolutional neural network (CNN) classification model from Kachuee et al.~\cite{kachuee_ecg_2018}.
We do not use data augmentation, as we are interested in seeing whether our visualizations can elucidate that certain classes are underrepresented.  
The model was trained for ten epochs on the training set (\textit{n} = 81,123), resulting in a final overall accuracy of 98.3\% on the test set (\textit{n} = 20,284).  
The breakdown of classes and performances on each is in Table 1.

\begin{table}[h]
    \centering
    \begin{tabular}{c|c|c}
        Class & \% of Examples & Test Set Accuracy\\
        \toprule
        Normal & 89.3\% & 99.6\% \\ \hline 
        Supraventricular Ectopic & 2.7\% & 70.5\% \\ \hline
        Ventricular Ectopic & 7.1\% & 95.7\% \\\hline
        Fusion & 0.8\% & 70.4\% \\\hline
        Overall & -- & 98.3\% \\
        \bottomrule
    \end{tabular}
    \caption{Classes used in the ECG beat classification task, along with their distribution in the dataset and the model’s test set performance.}
    \label{tab:my_label}
\end{table}

\subsection{Grounding Model Output in K-Nearest Neighbors}\label{sec:knn}
\begin{figure}
    \centering
    \includegraphics[width=0.86\linewidth]{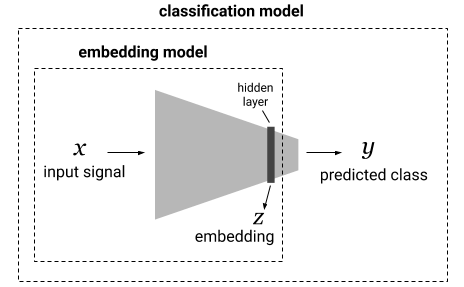}
    \caption{To compute nearest neighbors, we extract an embedding model from the original classification model, where the output is a learned representation (i.e., the activation of a hidden layer).  We use it to embed the training data examples and rank them by similarity to the input in this learned embedding space, returning the most similar. }
    \label{fig:embedding_model}
\end{figure}

The KNN module displays the model’s output for a particular example in terms of its nearest neighbors in the data. The nearest neighbors are computed similarly to prior work \cite{papernot_deep_2018,shin1999memory,caruana1999case}: Given a neural network model trained to perform the classification task (the \textit{classification model}), we first define an \textit{embedding model}, whose output is the activations of one of the model’s hidden layers (see Figure \ref{fig:embedding_model}).  We use this to embed all the training examples.  Then, for a given new input example, we embed it and then use KNN to find the most similar training examples in this learned representation space.  

Computing nearest neighbors in the learned embedding space of the classification model provides the advantage of harnessing the classification model’s representational capacity.  As this learned space encodes higher level features relevant to the task, these features are taken into account when calculating similar examples.
This step is particularly important to our goal of  narrowing the \textit{gulf of evaluation} \cite{hutchins1985direct} as it provides a way for users to understand the model’s output in terms of higher-level concepts that align with how they think about the task.
The model output can then be visualized in terms of the nearest neighbors. 

Different visual components display the nearest neighbors at varying levels of granularity, which together address our design goals G1, G2 and G3. They include an aggregate view of the neighbors’ class labels, a unit visualization of individual neighbors that encodes their class and distance from the input, and an overlaid display of the raw input examples associated with each neighbor.

\begin{figure}[]
    \centering
    \begin{subfigure}[t]{\linewidth}
         \centering
         \includegraphics[width=\textwidth]{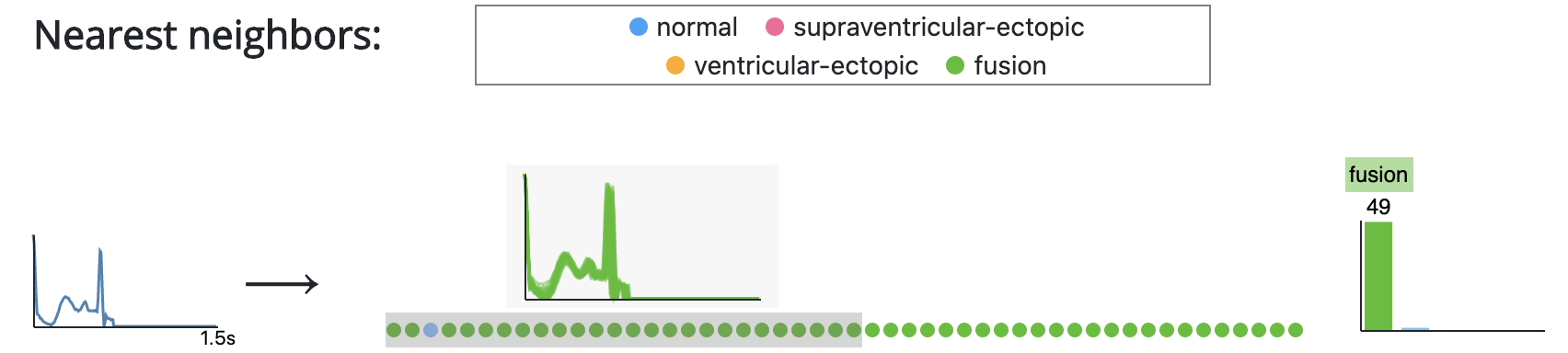}
         \caption{}
         \label{fig:consistency}
     \end{subfigure}
     \hfill
     \begin{subfigure}[t]{\linewidth}
         \centering
         \includegraphics[width=\textwidth]{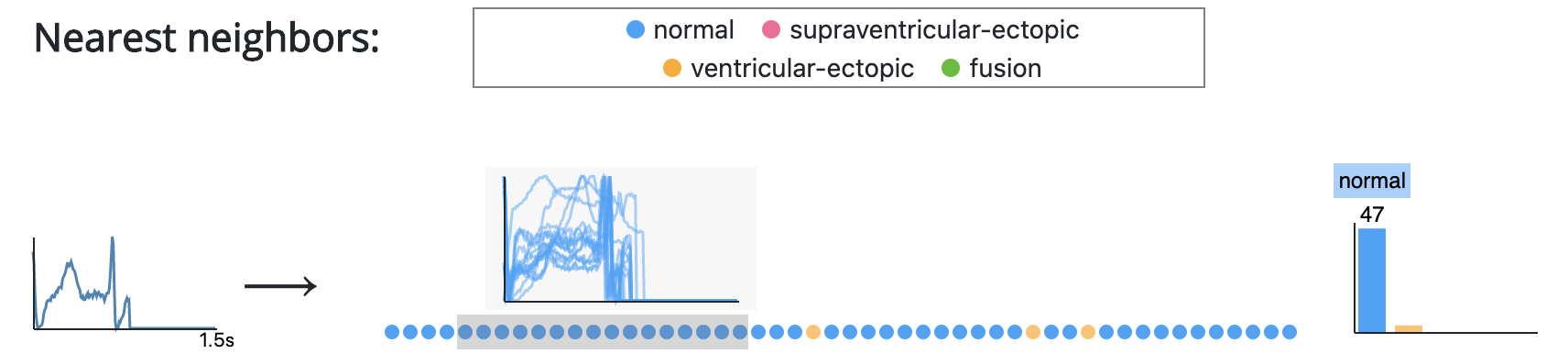}
         \caption{}
         \label{fig:variance}
     \end{subfigure}
    
    \caption{Examples of the KNN module. On the left is the input signal, and on the right is a histogram of class labels for the 50 nearest neighbors. In the center, each dot represents an individual nearest neighbor, ordered by similarity to the input. The plot above overlays the signals in the selected region. (a) shows an example where the neighbors are very consistent, and (b) shows an example where they are much noisier.}
    \label{fig:reliability}
\end{figure}

\textbf{ECG Case Study.} For the ECG beat classification task, we use the CNN classification model described in Sec. \ref{sec:case_study}, and we define the embedding model to output the activations from the final hidden layer (a 32-dimensional vector). We use Euclidean distance in this space to rank the embeddings of the training examples by their similarity to a particular input. We retrieve the 50 nearest neighbors for visualization. 

Figure \ref{fig:reliability} shows example ECG beats in the interface.  Throughout the interface, color encodes class labels (e.g., orange waveforms, dots, and bars correspond to ventricular ectopic examples).  The aggregate view is a histogram of class labels present in the nearest neighbors, ordered by class frequency to identify the majority class and distribution of other classes. The exact count of each class appears on hover for each bar in the histogram. The unit visualization of individual neighbors is a series of dots arrayed horizontally and ordered by similarity to the input.  Users can see, for example, within the nearest neighbors if certain classes are more similar to the input.  When prototyping this component, we also considered designs that encoded the absolute similarity (e.g., placing two neighbors that were more similar nearer to each other).  However, we decided against this, since the absolute similarity (i.e., Euclidean distance in the learned embedding space) is not a value that is meaningful or familiar to the user.  Additionally, the distribution of these values is more complicated to visualize, since the distances between neighbors are inconsistent.  In our prototypes, for example, there were often clusters of points that densely overlapped and did not facilitate selecting and viewing individual examples.  

To visualize the raw input examples, users can brush over specific segments of the ordered dots.  The brush is initialized to the first five neighbors, since these represent the most similar examples.  Because the ECG data is signal-based, we choose to visualize the neighbors by overlapping signals on a single plot that appears above the brush.  This allows users to visually assess consistency amongst the neighbors\,---\,for example, if the neighbors are very consistent, the overlaid plot will look very similar to a single signal. If they are more varied, the overlaid plot will appear comparably noisy.  Outliers are also visible, since they appear as a distinct waveform that does not follow in the same pattern as the other signals. By moving and adjusting the brush to cover specific segments of the neighbors, users can home in on and compare examples from specific classes or individual outliers.



\subsection{Interactively Editing Model Inputs}\label{sec:editor}

To address our final design goal (G4), the editor module allows users to apply transformations to the input and re-run the modified input through the model to see how the output changes.  For example, users can apply transformations that they expect to be class-preserving and check whether the model’s output changes drastically.  

The available transformations should help narrow the gulf of execution in the interface by providing transformations that align with users' existing ways of thinking about the data and task.  For example, in a dataset of natural images, it does not make sense to invert the colors because that is not something that would occur naturally, and does not reflect thought processes of people analyzing images.  We also would not want to provide transformations like editing individual pixels, which operate at a much lower-level than a person looking at an image would consider.  To come up with transformations that are data-specific (meaning they reflect how users think about modifying a specific type of data, like images or ECG signals), relevant to the task (meaning they reflect higher-level factors that users consider important to the task at hand), and aligned with the target users’ level of understanding, we emphasize the importance of working with domain experts and other intended end users to design them.

\begin{figure}[h]
    \centering
    \includegraphics[width=0.85\linewidth]{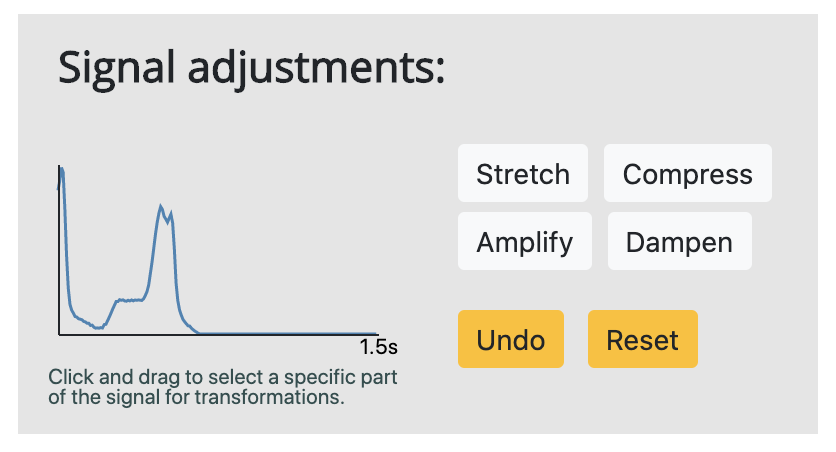}
    \caption{The editing toolbar allows users to apply specific transformations or combinations of transformations to the input signal.  The transformations can be applied to the entire signal, or to a specific user-selected region.   This allows users to select and transform clinically-meaningful segments of the signal (e.g., “stretch the QRS complex”).}
    \label{fig:editor1}
\end{figure}

\textbf{ECG Case Study.} 
For the ECG beat classification task, the editor consists of four transformations which we arrived at through discussion with a cardiologist: amplify, dampen, stretch, and compress.  These transformations can be applied to the entire input signal, or to specific user-defined regions using the brushing functionality. Together, they allow for a large space of possible adjustments to the input signal.  There are other options that could be explored here, such as automatically detecting certain important sections of the signal (e.g., “P wave” or “QRS complex”) to transform instead of having users select them themselves.  

\begin{figure}
    \centering
    \includegraphics[width=\linewidth]{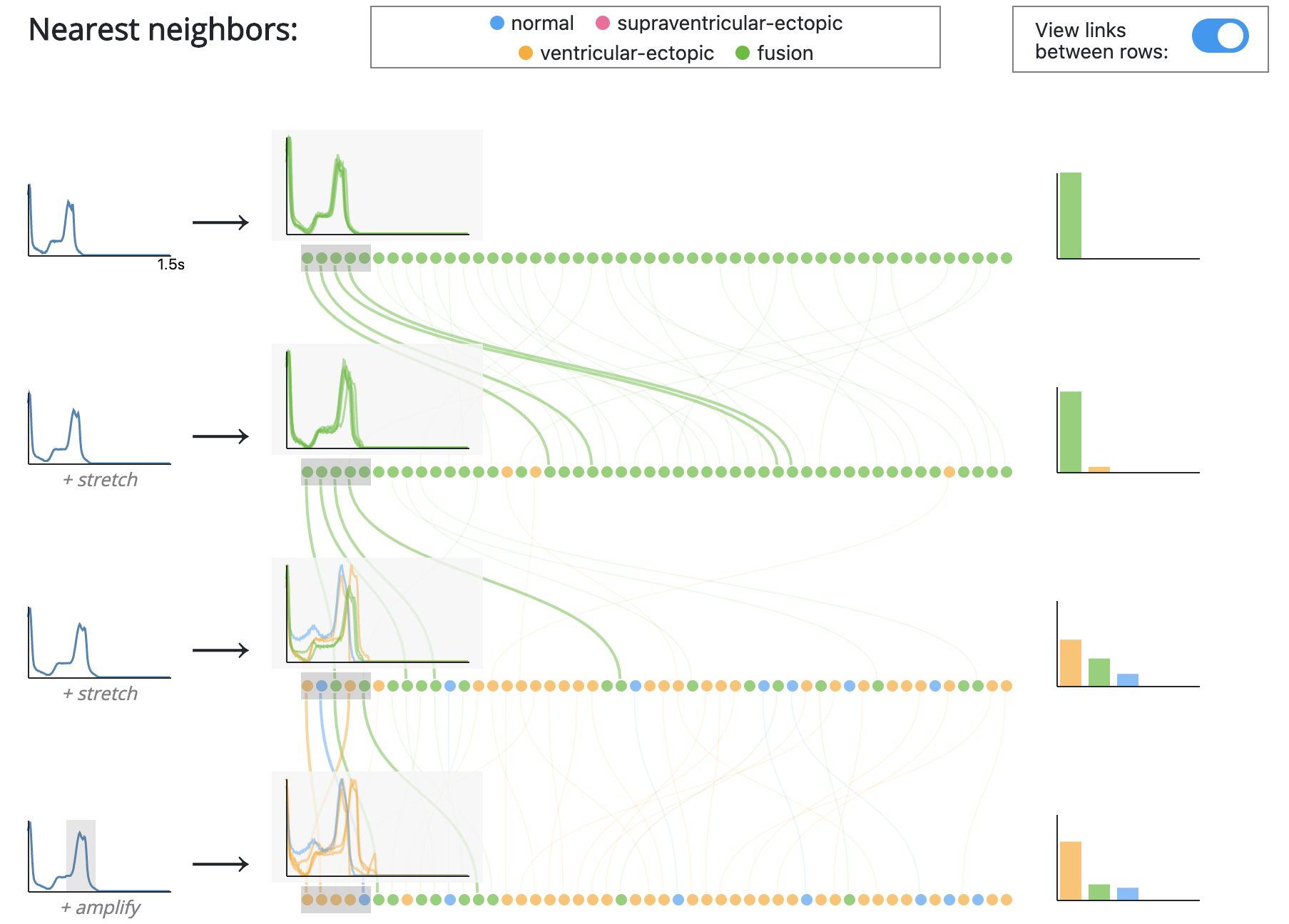}
    \caption{As transformations are applied, new rows appear with the transformed input and corresponding output.  Links between each row indicate neighbors that are shared.  Links originating from a row’s selection are more visible, while the rest are more transparent. Users can get a general sense of how much the nearest neighbors change (by assessing the overall density of links) as well as the specific movements of particular neighbors or sets of neighbors.}
    \vspace{-3mm}
    \label{fig:editor2}
\end{figure}

Once the transformation has been applied, a new row appears below the original output, displaying the new output.  The color encoding as well as highlighting on hover enables tracing how the class distribution changes overall, while links between neighbors that are shared across rows enables tracking how individual examples shift in similarity.  The editing toolbar is pictured in Figure \ref{fig:editor1}, and an example of the output after several transformations is in Figure \ref{fig:editor2}. 


\subsection{Enabling an Integrated Visual Analysis Workflow}\label{sec:use_cases}


Here, using the ECG case study, we expand upon several specific ways that a user can interact the visual analytics modules to assess the model's reliability, understand why it is uncertain, and check whether its reasoning aligns with domain knowledge: 

\subsubsection{Assessing consistency among nearest neighbors to understand prediction reliability and data limitations}
Users can assess the reliability of the prediction in multiple ways.  First, the aggregate distribution of class labels can convey the model’s uncertainty in the prediction (i.e., the majority class label).  For example, if 45 neighbors are normal, this conveys more certainty about the prediction than if only 25 neighbors are normal, and the rest are spread out across other classes.  

Second, by viewing the class labels of the unit visualization representing individual neighbors, users can see how similar the neighbors from non-majority classes are to neighbors from the majority class.  For example, if there are 40 neighbors labeled normal and 10 neighbors labeled fusion, are those 10 the most similar to the input?  Or do they appear closer to the latter end of the nearest neighbors?  If the neighbors from the non-majority class are the 10 most similar, this might indicate further unreliability of the ‘normal’ prediction. 

Third, visualizing the variance or consistency amongst the waveforms themselves can give insight into whether the input example is well-represented in the training data and whether the model is picking up on sensible high-level features common in the neighbors.  For example, if the overlaid plot of nearest neighbors shows examples that are very consistent and similar to the input in semantically meaningful ways (see Figure \ref{fig:consistency} for an example), it implies that the input is well-represented in the training data and that the model is picking up on the right concepts for this input. On the other hand, if the plot of nearest neighbor signals shows examples that are non-overlapping or not similar to the input (see Figure \ref{fig:variance} for an example), it implies that either examples like the input are not well-represented in the training data, or that the model is not learning the right features and therefore not finding those similar examples.

\subsubsection{Investigating neighbors from non-majority classes to characterize prediction uncertainty}

Typically, a classification model outputs a probability score indicating its certainty in its prediction.  Probability scores can alert the user to some uncertainty in the model, but they don’t give the user any additional information to understand \textit{why} the model is uncertain.  

\begin{figure}[h]
    \centering
    \begin{subfigure}[b]{\linewidth}
         \centering
         \includegraphics[width=\textwidth]{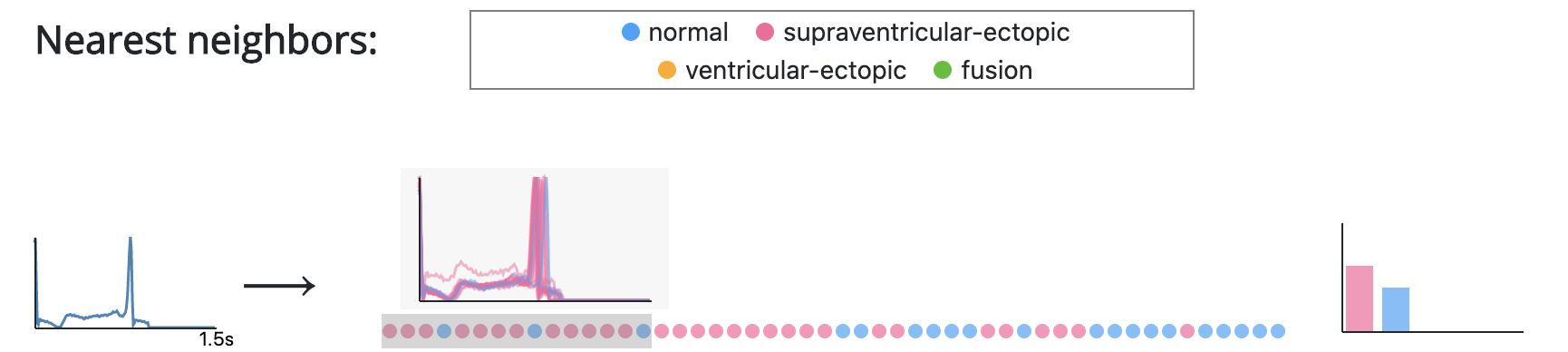}
         \caption{}
         \label{fig:uncertainty1}
     \end{subfigure}
     \hfill
     \begin{subfigure}[b]{\linewidth}
         \centering
         \includegraphics[width=\textwidth]{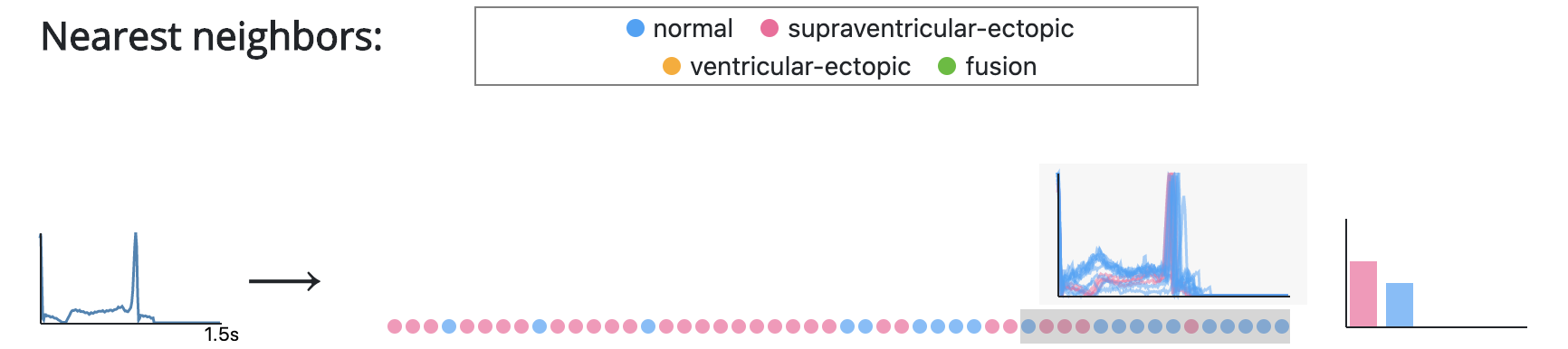}
         \caption{}
         \label{fig:uncertainty2}
     \end{subfigure}
    
    \caption{The user can home in on different examples to better understand the model's uncertainty. The view of the first 15 neighbors in (a) suggests that some of the model’s uncertainty is arising from the fact that normal beats can look very similar to supraventricular beats. Viewing the normal neighbors in (b) suggests that another reason for uncertainty is ambiguity around whether the input has a significant T-wave (the spike at the beginning of the signal).
 }
    \label{fig:uncertainty}
\end{figure}

In the KNN module, one way the model’s certainty is conveyed is through the aggregate distribution of class labels.  Beyond this, though, the user can further investigate why the model is uncertain by viewing and comparing examples from non-majority classes.  Brushing over specific selections of dots representing individual neighbors allows the user to better compare neighbors from different classes.  Take the example in Figure \ref{fig:uncertainty}: 30 of the neighbors have the class label supraventricular ectopic, and 20 have the label normal (these counts are visible upon hover in the aggregate histogram). In Figure \ref{fig:uncertainty1}, brushing over the first 15 neighbors reveals that most of them follow the same general pattern, and it looks very similar to the input.  The 3 normal neighbors in this selection also seem to follow this pattern\,---\,so some of the model’s uncertainty is arising from the fact that in the training data, there are normal beats that can look very similar to supraventricular beats.  In Figure \ref{fig:uncertainty2}, brushing over the last 15 neighbors reveals that most of them follow the same general pattern, but have a more elevated T-wave (the spike at the beginning of the signal) than the supraventricular ectopic neighbors.  A user might reason, then, that the model is split between supraventricular and normal, and one of the factors driving the uncertainty is whether or not the input has a significant T-wave.  

They could then use their domain knowledge to reason about how to proceed.  In this example, they might examine the input and decide that the T-wave is significantly depressed, making the input more similar to the supraventricular ectopic examples, and more confidently proceed with supraventricular ectopic as the correct class. Or, they might decide that the different classes present in the neighbors reflect legitimate ambiguities about what the correct beat type is, and choose to consult a second option or run additional tests.

\subsubsection{Comparing examples and labels against domain expectations to prompt critical questioning around the data}
If neighboring examples or their labels do not align with the user’s expectations, it can prompt questions from the user about the details of the data and how it was collected or labeled, areas that are too often not engaged with after a model’s deployment.  Crucially, seeing the signals themselves facilitates this type of critical thinking for people who are likely more familiar with the data and what it should look like than more abstract representations like feature weights.

\begin{figure}[h]
    \centering
    \includegraphics[width=\linewidth]{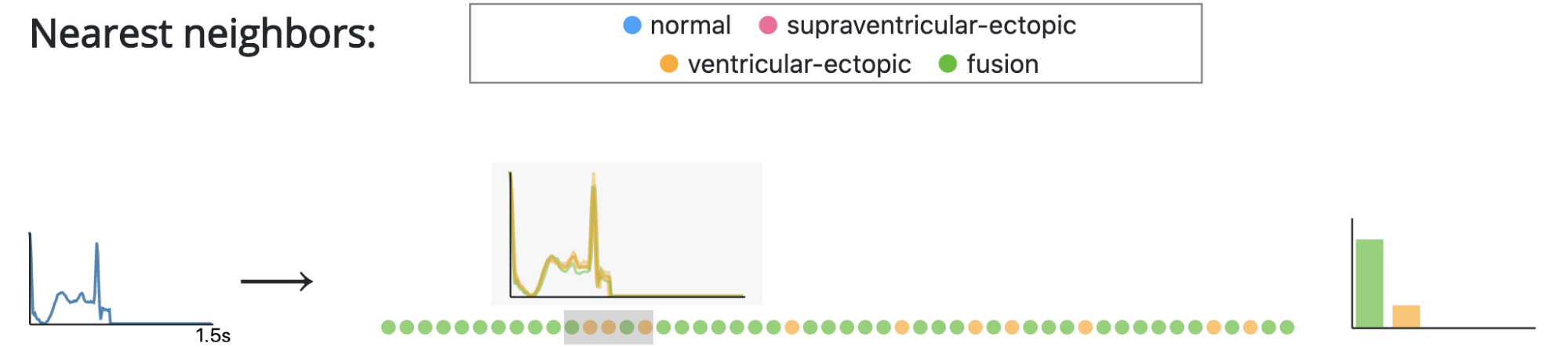}
    \caption{An example of neighbors that look similar but have different labels, due to a discrepancy in the additional information available during annotation versus at test-time.  Alerting users to such cases through viewing nearest neighbors can help prompt questions about the data, the annotation process, and limitations of the model.}
    \vspace{-5mm}
    \label{fig:fusion_example}
\end{figure}

In the ECG case study, for example, the data was annotated by physicians who had access to additional information about the beats preceding and following the input.  As a result, there are some examples in the dataset that look extremely similar but are labelled differently (presumably because the difference in their label was due to the information available during annotation that the model does not see).  In some cases, this leads to nearest neighbors that have different classes but look very similar (see Figure \ref{fig:fusion_example}).  Viewing the neighbors for a particular example can prompt questions about how the data was annotated and the subsequent limitations of the model, which would likely not arise if users were not able to view and compare specific similar examples. 

\subsubsection{Applying input transformations to check if model reasoning aligns with domain knowledge}
\begin{figure}[t]
    \centering
    \includegraphics[width=\linewidth]{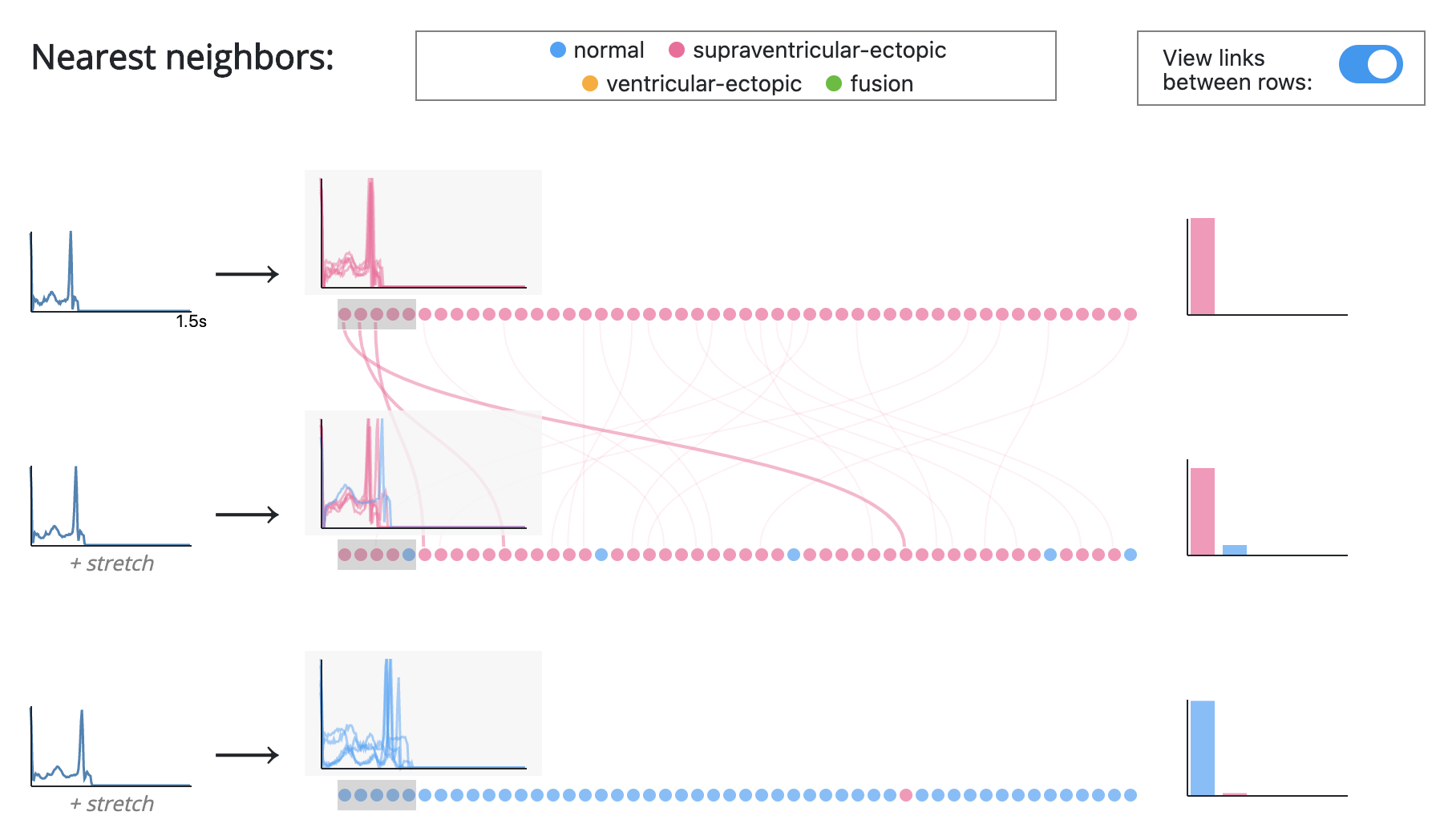}
    \caption{An example of using the editor to check if the model's reasoning aligns with domain expectations (i.e., stretching out a supraventricular ectopic beat should shift the prediction towards normal).}
    \label{fig:domain_knowledge}
\end{figure}

Checking if the model’s reasoning aligns with prior expectations of domain experts is important for building trust, especially in the clinical domain \cite{tonekaboni2019clinicians,cai2019hello}.  The editor module allows users to form hypotheses about how particular transformations should change the model’s output, and build confidence and intuition around the model’s reasoning by seeing if these hypotheses hold.  For example, the beat in Figure \ref{fig:domain_knowledge} is initially classified as supraventricular ectopic.  The user might hypothesize that since one indicator of supraventricular ectopic beats is narrowness, and this particular beat is narrow, that this is what the model is picking up on. Therefore, stretching the beat should change the model’s output, making it lean more towards normal. The user can apply this transformation in the editor to test their hypothesis. In this case, the model’s output does change to reflect more normal neighbors, confirming both the original hypothesis and that the model's behavior aligns with the user’s expectations from a clinical perspective. 

\subsubsection{Applying transformations to assess the model's sensitivity to small perturbations}
Aside from specific hypotheses about how a particular series of transformations should change the output, a user can still gauge the reliability of a particular prediction by performing ad hoc sensitivity analyses.  If the output changes drastically when the input is slightly tweaked, this can alert users to the fact that the prediction is precarious and encourage them not to be overly reliant on it.  On the other hand, if the output is relatively stable, this can be an additional indicator of model reliability.  

\section{Evaluative Studies with Medical Professionals}

To understand how effectively our visual analytics modules help users build intuition for ML model reliability, we evaluated our ECG beat classification case study with 14 participants recruited through our personal and professional networks: 3 fourth year medical students (P1-P3) and 11 physicians (P4-P14). The studies were certified by our institution as exempt from IRB review under Category 3.





\subsection{Study Design}

\begin{figure}[b]
    \centering
    \includegraphics[width=\linewidth]{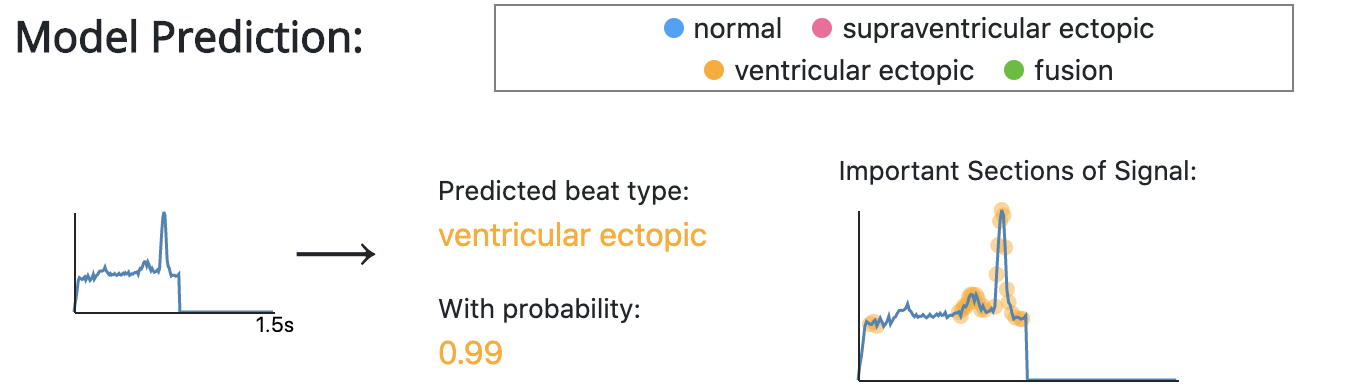}
    \caption{The baseline visualization consists of the predicted beat class, the probability with which that class was predicted, and highlighted segments of the beat considered most important for the prediction.}
    \label{fig:saliency}
\end{figure}

In order to study the effect of each of our modules independently, each participant experienced three conditions. The first two conditions were randomly ordered between our KNN visualization (without the editor) or a baseline feature-importance visualization, to understand the impact of example-based explanations on building intuition about the ML model.
To understand the impact of interactively editing inputs, participants experienced a third condition featuring the KNN visualization \textit{with} the input editor.
We chose to use feature importance as our baseline as it is a widely researched alternative to example-based explanations~\cite{du_techniques_2019,bhatt_explainable_2020}. 
The baseline condition, shown in Figure \ref{fig:saliency}, emulates the design of our KNN visualization, and feature importance is calculated via LIME~\cite{ribeiro_why_2016}, a commonly-used open-source method. 
In particular, LIME results are shown as highlighted regions that overlay the waveform, in line with existing approaches for visualizing ECG feature importance~\cite{mousavi2020han,tison2019automated}. 
We plot the feature importance values that are both above the 80th percentile and part of a continuous segment of neighboring important features, to better align with physicians’ existing ways of thinking about regions of an ECG signal.

Each condition was pre-populated with 12 input beats chosen from the test set and equally distributed among the four classes.  
We select beats such that 30\% in each condition have incorrect predictions (for the baseline condition, the prediction is the class with highest probability; for the KNN condition, the prediction is the class that makes up the majority of nearest neighbors). These incorrect predictions were aligned with the model's actual performance (e.g., we did not include incorrect predictions for normal beats since there are very few of those in reality; we included more incorrect predictions for supraventricular ectopic since the model’s performance for that class is worse).  

All studies were conducted via video conferencing.  
Participants were informed that their participation was voluntary, that they could decline to continue at any point, and that their identities would remain anonymous in any research output. 
Audio and video was recorded with their consent, and the average study length was 52 minutes. 
Participants were compensated with a \$30 gift card. 

At the start of each study, participants were told which four categories of beats they would be working with including the more granular information about beat types included with the original dataset (e.g., there are multiple pathologies that fall under the umbrella of ``ventricular ectopic''). 
We described that they would see ECG beats one-by-one, along with output from a machine learning model that had high overall performance.
Participants were asked to imagine a scenario where their workplace had adopted such a tool for beat classification, and they were both trying to consider the model’s output to make the best decision about a particular beat, as well as get a general sense of how the model worked. 
We introduced each interface as using a separate model to mitigate participants carrying over preconceptions from prior conditions.
For each condition, participants were given a brief demo and were then sent a link to open the visual analytics interface on their computer and asked to share their screen.  
We prompted them to click through the beats and, for each one, think out loud about how they were coming to a decision about the beat’s class, how they were incorporating the model’s output, and whether their perceptions about the model changed.  
At the end of each condition, we debriefed participants with questions about their general impressions of the model's capabilities, the interface, and the strengths and weaknesses of both.

\subsection{Quantitative Results}
We recorded the percent of cases in which participants agreed with the model (versus when they disagreed or were not sure). For cases in which the prediction was correct, the agreement rate was similar across conditions; however, when the prediction was incorrect, we found that participants were less likely to go with the model's prediction when they were using the KNN interface, with or without the input editor (Table \ref{table:quant}). Often in these cases, they did not explicitly ``disagree'' with the model, but wanted additional information about the signal and/or patient before committing to an answer. We expand on how our interface prompted these additional considerations in the following section.

\renewcommand{\arraystretch}{1.5}
\begin{table}[ht]
\centering 
\begin{tabular}{p{2cm}|p{1.4cm}p{1.4cm}p{1.75cm}} 
\textit{Pred. Accuracy} & \textit{Baseline} & \textit{KNN} & \textit{KNN + Editor} \\  
\hline 
Correct & 0.64 (0.2) & 0.7 (0.16) & 0.67 (0.12) \\ \hline 
Incorrect & 0.73 (0.23) & 0.48 (0.27) & 0.5 (0.24) \\
\hline 
\end{tabular}
\caption{The mean agreement rate for correct predictions (8 per condition) and incorrect predictions (4 per condition). The standard deviation across participants is in parentheses.} 
\label{table:quant} 
\end{table}

\subsection{Qualitative Observations}

When using our tools, visualizations of neighboring signals allowed participants to reason about the model's output in terms of clinically-meaningful concepts, and examining variation in these signals helped participants to build intuition about prediction reliability.
By inspecting the class histogram, ordering of neighbors, and neighboring signals, participants were able to relate the model's uncertainty to relevant challenges of the task.
Finally, participants used the editor to confirm if the model’s reasoning was sensible and to guide decision-making.



\subsubsection{Nearest neighbors enable reasoning with clinically-relevant concepts}

Visualizing nearest neighbors enabled participants to reason about the model in terms of clinically-relevant concepts by generalizing and comparing across neighbors. They would often notice a particular morphology present in the neighbors that helped them understand the model’s behavior and whether it was clinically sensible. One participant, pointing to a pattern present in all the neighboring signals, said \textit{``Yeah, ventricular.  It’s this elevation and this space that’s making it think ventricular''} [P4]. Another described, \textit{``The model is right\,---\,with ventricular ectopic, the QRS spike should be broad, which is present in all the similar examples''} [P13]. Overall, ten participants [P1, P3, P4-P5, P7-9, P12-14] reasoned about the model using high-level clinically-relevant concepts that they observed in the neighbors, such as \textit{``depression in the signal''} [P13], \textit{``slope right after the P-wave''} [P7], \textit{``presence of a T-wave''} [P8], or \textit{``P-R interval''} [P5].

In some cases, participants were unsure why neighbors were considered similar, or disagreed with their class labels. For example, one participant said, \textit{``these [neighbors] are supposed to be are ventricular ectopic... I think they're normal. I don't know what to make of this [output]''} [P2]. Such cases may be partly due to the fact that annotators had access to additional information about surrounding beats during annotation that is not available in the current dataset. Without this information, it can sometimes be unclear why a beat has the class label that it does.  While the model's output was confusing in these cases, visualizing neighbors did prompt additional questions about the data and labeling process. For example, one participant asked, \textit{``Some of these normal ones look like they could be abnormal, so I’d want to know why they were called normal and what that was based on''} [P6]. Another further hypothesized, \textit{``Most likely this data was correctly annotated [...] but it’s not using all that information here''} [P2].  



In contrast, with the baseline condition, participants often had difficulty extracting higher-level, clinically-relevant concepts from the feature importance visualization.  For example, echoing a sentiment shared by many, one participant said, \textit{``I don’t see how these blue [highlighted] areas are super helpful here... what are they trying to get at?''} [P7]. Another participant, who struggled trying to connect the explanation to the predicted class, said \textit{``I don’t understand how they go from this [pointing at highlighted areas] to saying that there’s some aspect of a ventricular beat in there''}  [P12]. Some others had difficulty figuring out what about the highlighted section was important\,---\,for example, one participant asked, \textit{``Why is it highlighted here, is it looking at the height of this, is it looking at width?  And why only this part?''} [P1]. In some cases, the highlighted areas \textit{did} align with participants' expectations, connecting these sections back to the prediction was not straightforward.  One participant noted, for example, \textit{``Sometimes it was highlighting things I would also consider, but I still thought its prediction was wrong.  I don’t have any intuition on that.  I guess it’s finding some features. I would want to know what those features are, see whether they’re useful, if they have any intuitive correlation''} [P2]. 

\subsubsection{Visualizing variation helps assess prediction reliability}

All participants said that they did not place as much weight on the model’s prediction when there was a lot of variance in the overlaid signals. Participants felt more confident in their answers when the overlaid signals were very consistent and similar. They were also able to distinguish between variation that was acceptable given the task and domain (e.g., \textit{``This input isn't as picture perfect, so it makes sense that the model shows some variation in the overlaid examples''} [P4]) compared to variation that was an indicator of unreliability (e.g., \textit{``[The model’s output] isn't giving me much information right now. If I was given this result I wouldn't just listen to the machine, I would want additional information''} [P4]). 

When using the baseline condition, most participants only felt reassured when the predicted probability was very high and the prediction aligned with their own. 
When this was not the case, we observed that participants had trouble understanding how to incorporate the probability score.
As a result, they often rationalized incorrect predictions\,---\,even when it went against their initial instincts.  For example, one participant saw an abnormal beat, started to say it was abnormal, but then changed her mind after looking at the predicted class, which (incorrectly) was normal: \textit{``I don’t think this is normal... well actually seeing that the machine thinks normal... I guess it has a small QRS and the T-wave has a normal slope. Okay, I’ll put this in the normal category''} [P7].  
Seven participants [P2-4, P7, P9-11] went through similar processes of rationalizing an incorrect prediction after having expressed an inclination towards the correct class.  

Even when they did not rationalize an incorrect prediction, participants often struggled with building intuition about the probability score or highlighted sections.
For instance, one participant thought out loud, \textit{``I don't know, it seems high probability for a weird looking one like this.  And I don’t know if it makes sense what it’s looking at here and calling important.  I’m not confident about this''} [P1]. Similarly, another said \textit{``I’d say this is definitely supraventricular, but the model's not giving it a high probability.  I’m really not sure why that would be''} [P11].
Eight participants [P1-2, P5-7, P11, P13-14] expressed similar difficulties in reasoning about the reliability of the prediction in the baseline interface.

\subsubsection{Nearest neighbors help characterize uncertainty and incorporate it into decision-making}

In the KNN visualization, a wide distribution of nearest neighbors classes is one sign of model uncertainty. In such situations, participants consistently homed in on differences using the overlaid plot of waveforms and aligned these differences with clinical concepts.  For example, one participant viewed a beat where neighbors were split between supraventricular ectopic and normal, noting \textit{``For supraventricular ectopic one thing you look for is whether or not it has a P-wave.  It’s unclear in the input.  These [brushing over supraventricular ectopic examples] are probably saying it isn’t a P-wave.  And these [brushing over normal examples] have the P-wave so they’re probably saying that the input does also and that’s why it should be normal''} [P5].

Similarly, participants often connected the distribution of nearest neighbors to natural ambiguities in the task. For example, one participant noticed some ventricular ectopic beats present in a fusion beat's neighbors\,---\,\textit{``Given that fusion is itself a combination of ventricular ectopic and normal, it makes sense that there’s uncertainty here, and that there are some yellow [ventricular ectopic] ones that look similar''} [P8]. Rather than distrusting the model, the ability to contextualize its uncertainty helped participants rationalize and move forward with its output. For instance, regarding neighbors split across classes, another participant said \textit{``I would be exactly split like the model is between supraventricular and ventricular ectopic.  The fact that the model is also split between those two makes me feel better, and I would do further testing [in person] to differentiate which one it is''} [P4].  

\begin{figure}
    \centering
    \begin{subfigure}[b]{\linewidth}
         \centering
         \includegraphics[width=\textwidth]{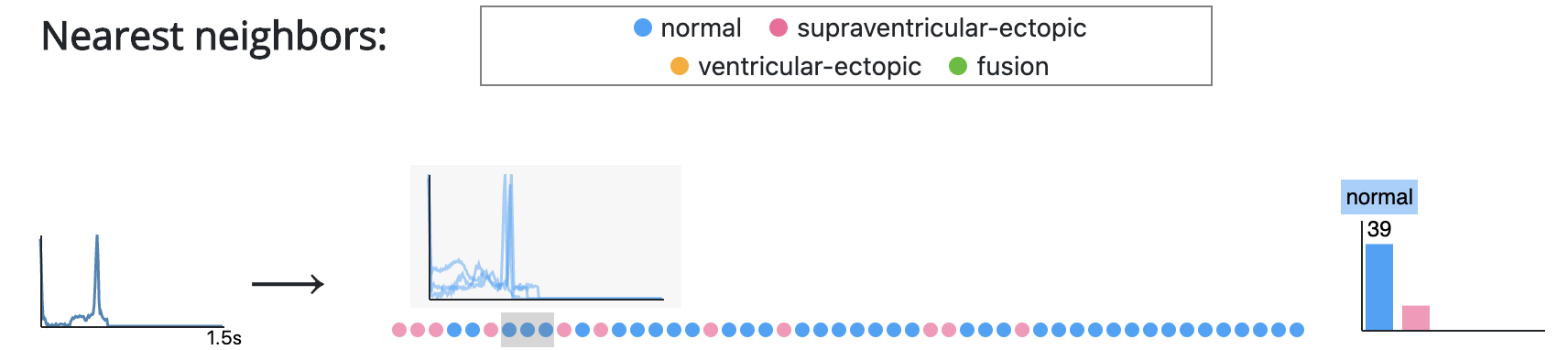}
         \caption{}
         \label{fig:study1}
     \end{subfigure}
     \hfill
     \begin{subfigure}[b]{\linewidth}
         \centering
         \includegraphics[width=\textwidth]{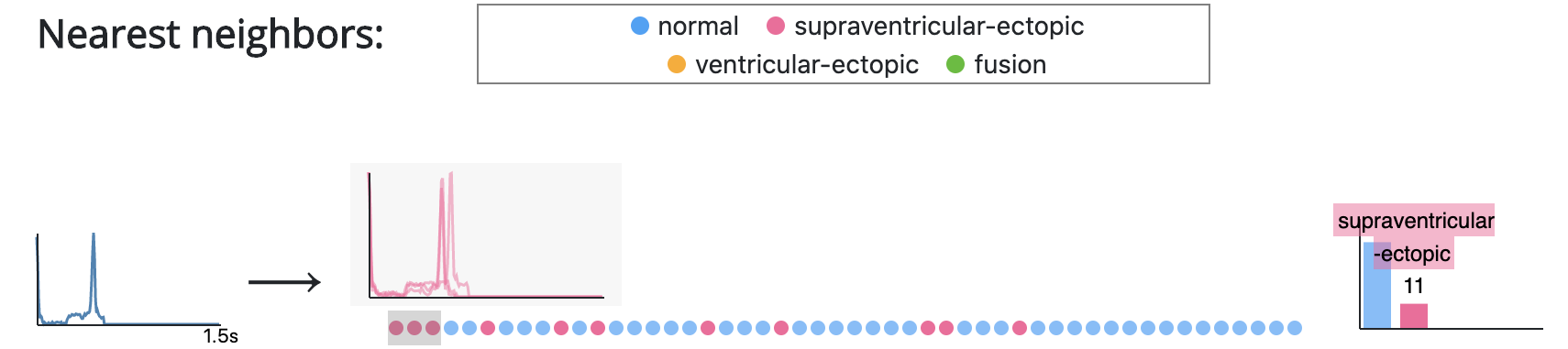}
         \caption{}
         \label{fig:study2}
     \end{subfigure}
    
    \caption{For this beat, one participant looked through some of the normal neighbors (a), comparing them to some of the supraventricular ectopic neighbors (b).  They reasoned that the normal examples, though they made up the majority of neighbors, were not more similar in clinically-meaningful ways to the input than the supraventricular ectopic examples.  As a result, they were able to arrive at the correct classification (supraventricular ectopic).}
    \label{fig:user1}
\end{figure}

Beyond making sense of the presence of multiple classes in the nearest neighbors, participants were also able use this information along with their domain knowledge during decision-making.  In many cases, upon viewing neighbors from the different classes, participants would realize that one of the classes was not actually similar to the input and, as a result, feel more confident in disregarding it.  For example, for the beat shown in Figure \ref{fig:user1}, one participant said \textit{``This is supraventricular ectopic.  [The model] is calling it normal, but the normal ones don’t look so similar.  The pink ones [supraventricular ectopic] look more like it because they also don't contain a P-wave''} [P14].  In other words, they were able to relate variation in the neighbors to clinical concepts (normal neighbors with a P-wave, supraventricular ectopic neighbors without), hypothesize why the model is uncertain (it isn’t sure whether the input example contains a P-wave), and use their own domain knowledge to determine how to proceed (the input does not actually have a P-wave, so go with supraventricular ectopic). Eight participants went through thought processes to better understand the model’s uncertainty and reconcile it with their knowledge of the domain knowledge [P4-8, P10, P13-14]. 

In contrast, when the model appeared less certain to those using the baseline (i.e., a lower probability score), participants had difficulty reasoning about why.  Many said they did not know why the probability was relatively low, or provided explanations based on their own knowledge as opposed to information from the feature importance visualization.  


\subsubsection{Editing inputs helps check model reasoning} 

\begin{figure}
    \centering
    \includegraphics[width=\linewidth]{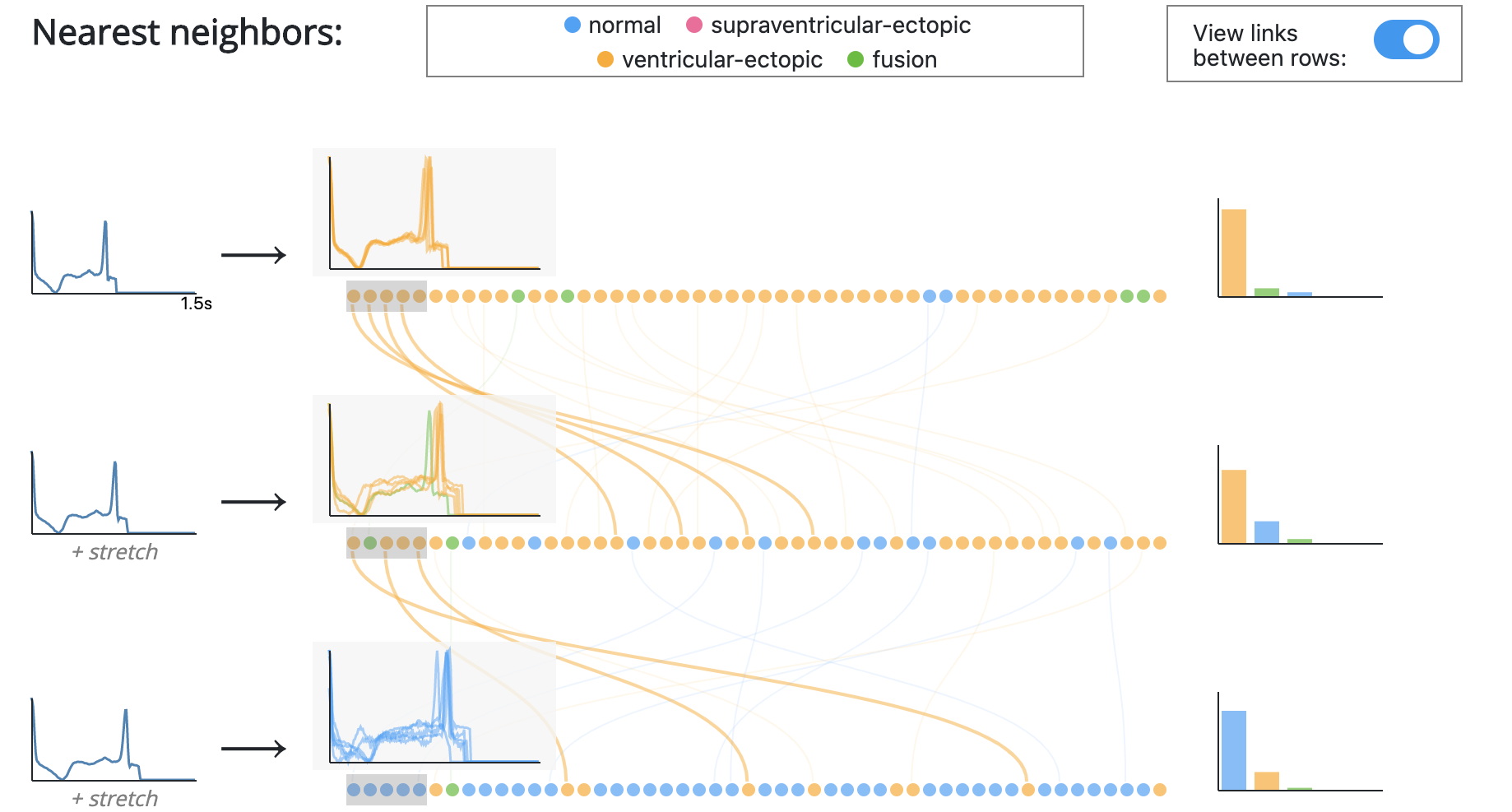}
    \caption{One participant hypothesized that the model was picking up on the narrowness of this beat in giving the prediction of ventricular ectopic, and thus stretching it would cause the neighbors to shift towards normal.  After applying the stretching transformation, and seeing that the nearest neighbors did change to be more normal, they felt more confident in the model’s reasoning for this beat and in classifying it as ventricular ectopic.}
    \label{fig:study3}
\end{figure}

Ten participants used the editor to formulate and test hypotheses about what would happen to the output after applying certain transformations [P4-9, P11-14].  They used this functionality as a way to ``sanity check'' the model's reasoning, and were more confident if it aligned with their expectations (and vice versa).  For example, one participant described using the editor to feel more confident in the model’s prediction for a beat (shown in Figure \ref{fig:study3}), which had mostly ventricular ectopic neighbors: \textit{``I'm not that confident with ventricular ectopic, and this looks almost normal. It’s a little narrow, which is partly what ventricular means, so I think that’s why this is saying ventricular and if I were to stretch it it would be normal. [Stretches the signal] And that’s exactly what happened.  That makes me more confident that this is more ventricular ectopic rather than normal.  Just because that’s exactly what my thought was and that’s exactly what happened when I did it''} [P9]. The same participant mentioned later on, \textit{``This is how I think of things.  If I can predict what’s going to happen I’m more likely to be confident in the decision.''}

Sometimes, however, participants applied a transformation but were not able to understand why the nearest neighbors changed as they did, or how to incorporate the observed change into downstream decision-making [P2, P4-5, P8, P10]. 
This situation typically occurred when the participant applied a transformation that they expected would shift the neighbors towards one of the non-normal beat classes, but instead skewed the neighbors towards normal\,---\,a behavior that reflects the model having learned less granular representations of beat classes that were under-represented in the data.
On one hand, this unexpected behavior prompted participants to rely less on the model's output in these cases\,---\,which, since the model is less accurate for these classes, is appropriate. 
At the same time, however, these instances were not able to offer participants useful insight into the model's reasoning. 

In other cases, participants applied several transformations separately to try and gauge the sensitivity of the prediction to small changes, as a way of assessing model reliability [P1, P3, P6-7, P10-11].  Sometimes several small transformations provided positive reinforcement\,---\,\textit{``Okay, this makes me more confident. When it’s normal, and then you do all these [transformations], I think it should mostly stay normal, which it is. It’s consistent so this all makes sense and I feel good with the machine''} [P1]. Other times, these transformations helped alert participants to the model's unreliability\,---\,\textit{``Seeing it switch so quickly from supraventricular ectopic to normal does affect my perception of whether it [the model] is good at telling those apart''} [P3]. 

With respect to the model's behavior more generally, some participants expressed an increased understanding in how the model worked after using the editor and observing what transformations tended to lead to a large change in the output.  One participant noted, \textit{``Doing these transformations is making me think about how this program works… I can tell that the narrowness of a beat affects the decision a lot for example''}  [P8]. 
Participants did not typically use the editor when the neighbors were consistent (both in terms of the shape of the signal and their class labels), because they did not feel the need to check the model's reasoning.  Other times, they chose not to use the editor because they could not think of a specific hypothesis they wanted to test\,---\,this was particularly true for the participants who were medical students, who often expressed that they ``didn't know enough'' but that someone with more experience might know what to test.  

\subsection{Study Limitations}

Several participants noted that the way the ECG beats were visualized was simplified.
For example, in practice, participants described that they would typically view a strip of beats from multiple leads, rather than one beat in isolation, and often with a grid overlaid to better measure distances. In some cases, this difference in display made participants more unsure about the class than they would have been if they had had their more familiar overlays. While the interface and task is simpler than it would be in a real clinical environment, in the current work our focus is more on developing and evaluating the proposed interpretability and visualization techniques, rather than developing a tool that could be deployed in a clinical setting (which would prompt an entirely different set of considerations).  

\section{Discussion and Future Work}

In this paper, we present two visual analytics modules to facilitate intuitive assessment of a machine learning model's reliability. 
Our work is motivated in part by interpretability needs elicited in prior work. For example, studies have found that communicating model limitations and uncertainty is important for building trust~\cite{cai2019hello,tonekaboni2019clinicians}, but that people have difficulty understanding the meaning of predicted probability scores and incorporating them into decision-making~\cite{bussone_role_2015}.
Other work has described the importance of users being able to ``sanity check'' a model's decision as a way to build trust~\cite{bhatt_explainable_2020,hong2020human,liao2020questioning}, but there have been few proposed methods or interfaces for doing so.
In response, our visual analytics modules are designed to allow users to interactively probe the model and to reason about its behavior through familiar examples grounded in their domain knowledge.
Users can explore a given input's nearest neighbors in the training data to better understand if and why the model is uncertain, and what high-level features the model is learning.  
They can further manipulate the input using domain-specific transformations to test hypotheses about the model's behavior or ensure that it is not overly sensitive to small changes.

Through think-aloud studies with 14 medical practitioners, we find that our visual analytics modules successfully achieve our design goals by helping participants reason about and interact with the model's output in ways that align with their existing conceptual models of the domain. 
Our studies demonstrate how grounding interpretability in real examples, facilitating comparison across them, and visualizing class distributions can help users grasp the model's uncertainty and connect it to relevant challenges of the task. 
Moreover, looking at and comparing real examples can also help users discover or ask questions about limitations of the data\,---\,and doing so does not damage trust, but can play an important role in building it. 

We also find that our interactive input editor, which offers semantically-meaningful and domain-specific transformations with which to probe the model, provides an effective way for users to sanity check the model's reasoning. 
Importantly, we find that participants in our study described the hypotheses they were testing in terms of higher-level features corresponding to their domain knowledge.  In contrast, the baseline\,---\,which implemented a commonly-used feature importance method~\cite{ribeiro_why_2016}\,---\,did not facilitate the same sorts of investigation. 
We found that this baseline interface demanded a large a mental leap from participants in order to understand how highlighted important sections of the waveform contributed to a high/low predicted probability.

At the same time, our results also point to limitations with the current design of our visual analytics interfaces and suggest opportunities for future work. 
We find that when the nearest neighbor waveforms looked significantly different than expected, participants had difficulty reasoning about why the model thought the neighbors were similar. 
We posit that part of participants' confusion was caused by the uneven distribution of beat classes in the training data, which affects the quality of nearest neighbors.
For example, supraventricular ectopic beats comprise only 2.7\% of training examples; thus, the model was neither able to precisely distinguish this beat from others, nor were there sufficient similar examples to fill the list of neighbors. 
However, this possibility of under-representation in the training data did not occur to participants when seeing low-quality neighbors. 
Aside from collecting sufficient data to compute better-quality neighbors, this result suggests the need for transparent communication of the model's training data distribution and its implications. 
If a user is then presented with output where the neighbors do not appear to make sense, they may be better equipped to understand why this might be the case. 
Indeed, we found that when we described this phenomena to participants after the conclusion of the study, they were able to understand why under-representation would affect the nearest neighbors\,---\,it had just not been on their radar previously. Similarly, Cai et al. \cite{cai2019hello} found the need for an \textit{``AI Primer''} for users to explain, in part, \textit{``AI-specific behavior that may be surprising.''} Our observations suggest specific use cases of and types of information to include in such a primer. 

In other cases, participants found it difficult to apply transformations using the input editor because the space of possible hypotheses was too open-ended.
Here, methods that generate counterfactual examples (i.e., similar example(s) that are classified differently)~\cite{wachter_counterfactual_2018,goyal_counterfactual_2019,mothilal2020explaining} might provide useful inspiration. 
These methods automatically generate modified inputs by finding small transformations that yield different predictions, but because they do not require any user intervention, they can return unrealistic examples that cannot be probed further. 
However, such methods could usefully bootstrap our input editor.
For example, automatically generated examples could help constrain the space of possible hypotheses to only those transformations that cause the greatest change in the model output.
Users could then bring their domain knowledge to bear on selecting semantically-meaningful examples to either visualize directly or as a starting point for further transformation. 

Finally, while we demonstrate our interface using an ECG case study, there is a significant opportunity for future work to investigate how these interface modules could be instantiated for other applications. Different data modalities will require different techniques to facilitate comparing examples and assessing variance\,---\,for example, while overlaying examples may be appropriate for signals and other image-based data, natural language data might require viewing examples separately and explicitly highlighting differences in wording. The right input modifications will also vary based on the application, and should emerge through working with the intended users. We imagine promising directions (e.g., building on Kim et al.~\cite{kim2018interpretability}) for allowing users to define and interact with meaningful high-level concepts in different types of data.


\section*{acknowledgements}
This research was sponsored by NSF Award \#1900991, and by the United States Air Force Research Laboratory under Cooperative Agreement Number FA8750-19-2-1000. The views and conclusions contained in this document are those of the authors and should not be interpreted as representing the official policies, either expressed or implied, of the United States Air Force or the U.S. Government. The U.S. Government is authorized to reproduce and distribute reprints for Government purposes notwithstanding any copyright notation herein.

\bibliographystyle{abbrv-doi}

\bibliography{bibliography}

\begin{thebibliography}{10}

\bibitem{aamodt_case-based_1994}
A.~Aamodt and E.~Plaza.
\newblock Case-{Based} {Reasoning}: {Foundational} {Issues}, {Methodological}
  {Variations}, and {System} {Approaches}.
\newblock {\em AI Communications}, 7(1):39--59, 1994. doi: {{%
10\hspace{.1pt}\discretionary{.}{%
}{.}\hspace{.4pt}3233\discretionary{/}{%
}{/}AIC\discretionary{%
}{-}{-}1994\discretionary{%
}{-}{-}7104}}


\bibitem{adhikari_leafage_2019}
A.~Adhikari, D.~M.~J. Tax, R.~Satta, and M.~Faeth.
\newblock {LEAFAGE}: {Example}-based and {Feature} importance-based
  {Explanations} for {Black}-box {ML} models.
\newblock In {\em 2019 {IEEE} {International} {Conference} on {Fuzzy} {Systems}
  ({FUZZ}-{IEEE})}, pp. 1--7. IEEE, New Orleans, LA, USA, June 2019. doi: {{%
10\hspace{.1pt}\discretionary{.}{%
}{.}\hspace{.4pt}1109\discretionary{/}{%
}{/}FUZZ\discretionary{%
}{-}{-}IEEE\hspace{.1pt}\discretionary{.}{%
}{.}\hspace{.4pt}2019\hspace{.1pt}\discretionary{.}{%
}{.}\hspace{.4pt}8858846}}


\bibitem{ajunwa2016hiring}
I.~Ajunwa.
\newblock The paradox of automation as anti-bias intervention.
\newblock {\em Forthcoming in Cardozo Law Review}, 2016.

\bibitem{barocas_hidden_2020}
S.~Barocas, A.~D. Selbst, and M.~Raghavan.
\newblock The hidden assumptions behind counterfactual explanations and
  principal reasons.
\newblock In {\em Proceedings of the 2020 {Conference} on {Fairness},
  {Accountability}, and {Transparency}}, pp. 80--89. ACM, Barcelona Spain, Jan.
  2020. doi: {{%
10\hspace{.1pt}\discretionary{.}{%
}{.}\hspace{.4pt}1145\discretionary{/}{%
}{/}3351095\hspace{.1pt}\discretionary{.}{%
}{.}\hspace{.4pt}3372830}}


\bibitem{basu_influence_2020}
S.~Basu, P.~Pope, and S.~Feizi.
\newblock Influence {Functions} in {Deep} {Learning} {Are} {Fragile}.
\newblock {\em arXiv:2006.14651 [cs, stat]}, June 2020.
\newblock arXiv: 2006.14651.

\bibitem{bhatt_explainable_2020}
U.~Bhatt, A.~Xiang, S.~Sharma, A.~Weller, A.~Taly, Y.~Jia, J.~Ghosh, R.~Puri,
  J.~M.~F. Moura, and P.~Eckersley.
\newblock Explainable machine learning in deployment.
\newblock In {\em Proceedings of the 2020 {Conference} on {Fairness},
  {Accountability}, and {Transparency}}, {FAT}* '20, pp. 648--657. Association
  for Computing Machinery, Barcelona, Spain, Jan. 2020. doi: {{%
10\hspace{.1pt}\discretionary{.}{%
}{.}\hspace{.4pt}1145\discretionary{/}{%
}{/}3351095\hspace{.1pt}\discretionary{.}{%
}{.}\hspace{.4pt}3375624}}


\bibitem{boggust2019embedding}
A.~Boggust, B.~Carter, and A.~Satyanarayan.
\newblock Embedding comparator: Visualizing differences in global structure and
  local neighborhoods via small multiples, 2019.

\bibitem{bucinca2020proxy}
Z.~Bu{\c{c}}inca, P.~Lin, K.~Z. Gajos, and E.~L. Glassman.
\newblock Proxy tasks and subjective measures can be misleading in evaluating
  explainable ai systems.
\newblock In {\em Proceedings of the 25th International Conference on
  Intelligent User Interfaces}, pp. 454--464, 2020.

\bibitem{bussone_role_2015}
A.~Bussone, S.~Stumpf, and D.~O'Sullivan.
\newblock The {Role} of {Explanations} on {Trust} and {Reliance} in {Clinical}
  {Decision} {Support} {Systems}.
\newblock In {\em 2015 {International} {Conference} on {Healthcare}
  {Informatics}}, pp. 160--169. IEEE, Dallas, TX, USA, Oct. 2015. doi: {{%
10\hspace{.1pt}\discretionary{.}{%
}{.}\hspace{.4pt}1109\discretionary{/}{%
}{/}ICHI\hspace{.1pt}\discretionary{.}{%
}{.}\hspace{.4pt}2015\hspace{.1pt}\discretionary{.}{%
}{.}\hspace{.4pt}26}}


\bibitem{cai_effects_2019}
C.~J. Cai, J.~Jongejan, and J.~Holbrook.
\newblock The effects of example-based explanations in a machine learning
  interface.
\newblock In {\em Proceedings of the 24th {International} {Conference} on
  {Intelligent} {User} {Interfaces}}, pp. 258--262. ACM, Marina del Ray
  California, Mar. 2019. doi: {{%
10\hspace{.1pt}\discretionary{.}{%
}{.}\hspace{.4pt}1145\discretionary{/}{%
}{/}3301275\hspace{.1pt}\discretionary{.}{%
}{.}\hspace{.4pt}3302289}}


\bibitem{cai2019human}
C.~J. Cai, E.~Reif, N.~Hegde, J.~Hipp, B.~Kim, D.~Smilkov, M.~Wattenberg,
  F.~Viegas, G.~S. Corrado, M.~C. Stumpe, et~al.
\newblock Human-centered tools for coping with imperfect algorithms during
  medical decision-making.
\newblock In {\em Proceedings of the 2019 CHI Conference on Human Factors in
  Computing Systems}, pp. 1--14, 2019.

\bibitem{cai2019hello}
C.~J. Cai, S.~Winter, D.~Steiner, L.~Wilcox, and M.~Terry.
\newblock ``hello ai'': Uncovering the onboarding needs of medical
  practitioners for human-ai collaborative decision-making.
\newblock {\em Proceedings of the ACM on Human-computer Interaction},
  3(CSCW):1--24, 2019.

\bibitem{carter_exploring_2019}
S.~Carter, Z.~Armstrong, L.~Schubert, I.~Johnson, and C.~Olah.
\newblock Exploring {Neural} {Networks} with {Activation} {Atlases}.
\newblock {\em Distill}, 4(3):10.23915/distill.00015, Mar. 2019. doi: {{%
10\hspace{.1pt}\discretionary{.}{%
}{.}\hspace{.4pt}23915\discretionary{/}{%
}{/}distill\hspace{.1pt}\discretionary{.}{%
}{.}\hspace{.4pt}00015}}


\bibitem{caruana1999case}
R.~Caruana, H.~Kangarloo, J.~Dionisio, U.~Sinha, and D.~Johnson.
\newblock Case-based explanation of non-case-based learning methods.
\newblock In {\em Proceedings of the AMIA Symposium}, p. 212. American Medical
  Informatics Association, 1999.

\bibitem{carvalho_machine_2019}
D.~V. Carvalho, E.~M. Pereira, and J.~S. Cardoso.
\newblock Machine {Learning} {Interpretability}: {A} {Survey} on {Methods} and
  {Metrics}.
\newblock {\em Electronics}, 8(8):832, July 2019. doi: {{%
10\hspace{.1pt}\discretionary{.}{%
}{.}\hspace{.4pt}3390\discretionary{/}{%
}{/}electronics8080832}}


\bibitem{doshi-velez_towards_2017}
F.~Doshi-Velez and B.~Kim.
\newblock Towards {A} {Rigorous} {Science} of {Interpretable} {Machine}
  {Learning}.
\newblock {\em arXiv:1702.08608 [cs, stat]}, Mar. 2017.
\newblock arXiv: 1702.08608.

\bibitem{du_techniques_2019}
M.~Du, N.~Liu, and X.~Hu.
\newblock Techniques for interpretable machine learning.
\newblock {\em Communications of the ACM}, 63(1):68--77, Dec. 2019. doi: {{%
10\hspace{.1pt}\discretionary{.}{%
}{.}\hspace{.4pt}1145\discretionary{/}{%
}{/}3359786}}


\bibitem{kaggle_dataset}
S.~Fazeli.
\newblock {\em ECG Heartbeat Categorization Dataset}.

\bibitem{gaube2021ai}
S.~Gaube, H.~Suresh, M.~Raue, A.~Merritt, S.~J. Berkowitz, E.~Lermer, J.~F.
  Coughlin, J.~V. Guttag, E.~Colak, and M.~Ghassemi.
\newblock Do as ai say: susceptibility in deployment of clinical decision-aids.
\newblock {\em NPJ digital medicine}, 4(1):1--8, 2021.

\bibitem{gilpin_explaining_2018}
L.~H. Gilpin, D.~Bau, B.~Z. Yuan, A.~Bajwa, M.~Specter, and L.~Kagal.
\newblock Explaining {Explanations}: {An} {Overview} of {Interpretability} of
  {Machine} {Learning}.
\newblock In {\em 2018 {IEEE} 5th {International} {Conference} on {Data}
  {Science} and {Advanced} {Analytics} ({DSAA})}, pp. 80--89. IEEE, Turin,
  Italy, Oct. 2018. doi: {{%
10\hspace{.1pt}\discretionary{.}{%
}{.}\hspace{.4pt}1109\discretionary{/}{%
}{/}DSAA\hspace{.1pt}\discretionary{.}{%
}{.}\hspace{.4pt}2018\hspace{.1pt}\discretionary{.}{%
}{.}\hspace{.4pt}00018}}


\bibitem{goyal_counterfactual_2019}
Y.~Goyal, Z.~Wu, J.~Ernst, D.~Batra, D.~Parikh, and S.~Lee.
\newblock Counterfactual {Visual} {Explanations}.
\newblock In {\em Proceedings of the 36th {International} {Conference} on
  {Machine} {Learning}}, vol.~97. Long Beach, California, USA, June 2019.
\newblock arXiv: 1904.07451.

\bibitem{heimerl2018interactive}
F.~Heimerl and M.~Gleicher.
\newblock Interactive analysis of word vector embeddings.
\newblock In {\em Computer Graphics Forum}, vol.~37, pp. 253--265. Wiley Online
  Library, 2018.

\bibitem{hirsch2017designing}
T.~Hirsch, K.~Merced, S.~Narayanan, Z.~E. Imel, and D.~C. Atkins.
\newblock Designing contestability: Interaction design, machine learning, and
  mental health.
\newblock In {\em Proceedings of the 2017 Conference on Designing Interactive
  Systems}, DIS '17, p. 95–99. Association for Computing Machinery, New York,
  NY, USA, 2017. doi: {{%
10\hspace{.1pt}\discretionary{.}{%
}{.}\hspace{.4pt}1145\discretionary{/}{%
}{/}3064663\hspace{.1pt}\discretionary{.}{%
}{.}\hspace{.4pt}3064703}}


\bibitem{hong_human_2020}
S.~R. Hong, J.~Hullman, and E.~Bertini.
\newblock Human {Factors} in {Model} {Interpretability}: {Industry}
  {Practices}, {Challenges}, and {Needs}.
\newblock {\em Proceedings of the ACM on Human-Computer Interaction},
  4(CSCW1):1--26, May 2020. doi: {{%
10\hspace{.1pt}\discretionary{.}{%
}{.}\hspace{.4pt}1145\discretionary{/}{%
}{/}3392878}}


\bibitem{hong2020human}
S.~R. Hong, J.~Hullman, and E.~Bertini.
\newblock Human factors in model interpretability: Industry practices,
  challenges, and needs.
\newblock {\em Proceedings of the ACM on Human-Computer Interaction},
  4(CSCW1):1--26, 2020.

\bibitem{hutchins1985direct}
E.~L. Hutchins, J.~D. Hollan, and D.~A. Norman.
\newblock Direct manipulation interfaces.
\newblock {\em Human--computer interaction}, 1(4):311--338, 1985.

\bibitem{jacovi2021formalizing}
A.~Jacovi, A.~Marasovi{\'c}, T.~Miller, and Y.~Goldberg.
\newblock Formalizing trust in artificial intelligence: Prerequisites, causes
  and goals of human trust in ai.
\newblock In {\em Proceedings of the 2021 ACM Conference on Fairness,
  Accountability, and Transparency}, pp. 624--635, 2021.

\bibitem{jesus2021can}
S.~Jesus, C.~Bel{\'e}m, V.~Balayan, J.~Bento, P.~Saleiro, P.~Bizarro, and
  J.~Gama.
\newblock How can i choose an explainer? an application-grounded evaluation of
  post-hoc explanations.
\newblock In {\em Proceedings of the 2021 ACM Conference on Fairness,
  Accountability, and Transparency}, pp. 805--815, 2021.

\bibitem{jiang2017artificial}
F.~Jiang, Y.~Jiang, H.~Zhi, Y.~Dong, H.~Li, S.~Ma, Y.~Wang, Q.~Dong, H.~Shen,
  and Y.~Wang.
\newblock Artificial intelligence in healthcare: past, present and future.
\newblock {\em Stroke and vascular neurology}, 2(4):230--243, 2017.

\bibitem{kachuee_ecg_2018}
M.~Kachuee, S.~Fazeli, and M.~Sarrafzadeh.
\newblock {ECG} {Heartbeat} {Classification}: {A} {Deep} {Transferable}
  {Representation}.
\newblock In {\em 2018 {IEEE} {International} {Conference} on {Healthcare}
  {Informatics} ({ICHI})}, pp. 443--444. IEEE, New York, NY, June 2018. doi:
  {{%
10\hspace{.1pt}\discretionary{.}{%
}{.}\hspace{.4pt}1109\discretionary{/}{%
}{/}ICHI\hspace{.1pt}\discretionary{.}{%
}{.}\hspace{.4pt}2018\hspace{.1pt}\discretionary{.}{%
}{.}\hspace{.4pt}00092}}


\bibitem{kim_interactive_2015}
B.~Kim.
\newblock {\em Interactive and {Interpretable} {Machine} {Learning} {Models}
  for {Human} {Machine} {Collaboration}}.
\newblock PhD thesis, Massachusetts Institute of Technology, Cambridge, MA,
  June 2015.

\bibitem{kim_examples_2016}
B.~Kim, R.~Khanna, and O.~O. Koyejo.
\newblock Examples are not enough, learn to criticize! {Criticism} for
  {Interpretability}.
\newblock In D.~D. Lee, M.~Sugiyama, U.~V. Luxburg, I.~Guyon, and R.~Garnett,
  eds., {\em Advances in {Neural} {Information} {Processing} {Systems} 29}, pp.
  2280--2288. Curran Associates, Inc., 2016.

\bibitem{kim2018interpretability}
B.~Kim, M.~Wattenberg, J.~Gilmer, C.~Cai, J.~Wexler, F.~Viegas, et~al.
\newblock Interpretability beyond feature attribution: Quantitative testing
  with concept activation vectors (tcav).
\newblock In {\em International conference on machine learning}, pp.
  2668--2677. PMLR, 2018.

\bibitem{koh_understanding_2017}
P.~W. Koh and P.~Liang.
\newblock Understanding {Black}-box {Predictions} via {Influence} {Functions}.
\newblock In {\em Proceedings of the 34th {International} {Conference} on
  {Machine} {Learning}}, vol.~70. Sydney, Australia, July 2017.
\newblock arXiv: 1703.04730.

\bibitem{kulesza_principles_2015}
T.~Kulesza, M.~Burnett, W.-K. Wong, and S.~Stumpf.
\newblock Principles of {Explanatory} {Debugging} to {Personalize}
  {Interactive} {Machine} {Learning}.
\newblock In {\em Proceedings of the 20th {International} {Conference} on
  {Intelligent} {User} {Interfaces} - {IUI} '15}, pp. 126--137. ACM Press,
  Atlanta, Georgia, USA, 2015. doi: {{%
10\hspace{.1pt}\discretionary{.}{%
}{.}\hspace{.4pt}1145\discretionary{/}{%
}{/}2678025\hspace{.1pt}\discretionary{.}{%
}{.}\hspace{.4pt}2701399}}


\bibitem{lai_human_2019}
V.~Lai and C.~Tan.
\newblock On {Human} {Predictions} with {Explanations} and {Predictions} of
  {Machine} {Learning} {Models}: {A} {Case} {Study} on {Deception} {Detection}.
\newblock In {\em Proceedings of the {Conference} on {Fairness},
  {Accountability}, and {Transparency} - {FAT}* '19}, pp. 29--38. ACM Press,
  Atlanta, GA, USA, 2019. doi: {{%
10\hspace{.1pt}\discretionary{.}{%
}{.}\hspace{.4pt}1145\discretionary{/}{%
}{/}3287560\hspace{.1pt}\discretionary{.}{%
}{.}\hspace{.4pt}3287590}}


\bibitem{lee2004trust}
J.~D. Lee and K.~A. See.
\newblock Trust in automation: Designing for appropriate reliance.
\newblock {\em Human factors}, 46(1):50--80, 2004.

\bibitem{liao2020questioning}
Q.~V. Liao, D.~Gruen, and S.~Miller.
\newblock Questioning the ai: Informing design practices for explainable ai
  user experiences.
\newblock In {\em Proceedings of the 2020 CHI Conference on Human Factors in
  Computing Systems}, pp. 1--15, 2020.

\bibitem{lipton1990contrastive}
P.~Lipton.
\newblock Contrastive explanation.
\newblock {\em Royal Institute of Philosophy Supplements}, 27:247--266, 1990.

\bibitem{liu2019latent}
Y.~Liu, E.~Jun, Q.~Li, and J.~Heer.
\newblock Latent space cartography: Visual analysis of vector space embeddings.
\newblock In {\em Computer Graphics Forum}, vol.~38, pp. 67--78. Wiley Online
  Library, 2019.

\bibitem{lundberg_unified_2017}
S.~Lundberg and S.-I. Lee.
\newblock A {Unified} {Approach} to {Interpreting} {Model} {Predictions}.
\newblock In {\em Advances in {Neural} {Information} {Processing} {Systems} 30
  ({NIPS} 2017)}, Nov. 2017.

\bibitem{michelini_multigrid_2019}
P.~N. Michelini, H.~Liu, and D.~Zhu.
\newblock Multigrid {Backprojection} {Super}–{Resolution} and {Deep} {Filter}
  {Visualization}.
\newblock {\em Proceedings of the AAAI Conference on Artificial Intelligence},
  33:4642--4650, July 2019. doi: {{%
10\hspace{.1pt}\discretionary{.}{%
}{.}\hspace{.4pt}1609\discretionary{/}{%
}{/}aaai\hspace{.1pt}\discretionary{.}{%
}{.}\hspace{.4pt}v33i01\hspace{.1pt}\discretionary{.}{%
}{.}\hspace{.4pt}33014642}}


\bibitem{miller1956magical}
G.~A. Miller.
\newblock The magical number seven, plus or minus two: Some limits on our
  capacity for processing information.
\newblock {\em Psychological review}, 63(2):81, 1956.

\bibitem{miller2019explanation}
T.~Miller.
\newblock Explanation in artificial intelligence: Insights from the social
  sciences.
\newblock {\em Artificial intelligence}, 267:1--38, 2019.

\bibitem{moody2001impact}
G.~B. Moody and R.~G. Mark.
\newblock The impact of the mit-bih arrhythmia database.
\newblock {\em IEEE Engineering in Medicine and Biology Magazine},
  20(3):45--50, 2001.

\bibitem{mothilal_explaining_2020}
R.~K. Mothilal, A.~Sharma, and C.~Tan.
\newblock Explaining machine learning classifiers through diverse
  counterfactual explanations.
\newblock In {\em Proceedings of the 2020 {Conference} on {Fairness},
  {Accountability}, and {Transparency}}, pp. 607--617. ACM, Barcelona Spain,
  Jan. 2020. doi: {{%
10\hspace{.1pt}\discretionary{.}{%
}{.}\hspace{.4pt}1145\discretionary{/}{%
}{/}3351095\hspace{.1pt}\discretionary{.}{%
}{.}\hspace{.4pt}3372850}}


\bibitem{mothilal2020explaining}
R.~K. Mothilal, A.~Sharma, and C.~Tan.
\newblock Explaining machine learning classifiers through diverse
  counterfactual explanations.
\newblock In {\em Proceedings of the 2020 Conference on Fairness,
  Accountability, and Transparency}, pp. 607--617, 2020.

\bibitem{mousavi2020han}
S.~Mousavi, F.~Afghah, and U.~R. Acharya.
\newblock Han-ecg: An interpretable atrial fibrillation detection model using
  hierarchical attention networks.
\newblock {\em arXiv preprint arXiv:2002.05262}, 2020.

\bibitem{mulligan2019shaping}
D.~K. Mulligan, D.~Kluttz, and N.~Kohli.
\newblock Shaping our tools: Contestability as a means to promote responsible
  algorithmic decision making in the professions.
\newblock {\em Available at SSRN 3311894}, 2019.

\bibitem{nunes2017review}
I.~Nunes and D.~Jannach.
\newblock A systematic review and taxonomy of explanations in decision support
  and recommender systems.
\newblock {\em User Modeling and User-Adapted Interaction},
  27(3–5):393–444, Dec. 2017. doi: {{%
10\hspace{.1pt}\discretionary{.}{%
}{.}\hspace{.4pt}1007\discretionary{/}{%
}{/}s11257\discretionary{%
}{-}{-}017\discretionary{%
}{-}{-}9195\discretionary{%
}{-}{-}0}}


\bibitem{papernot_deep_2018}
N.~Papernot and P.~McDaniel.
\newblock Deep k-{Nearest} {Neighbors}: {Towards} {Confident}, {Interpretable}
  and {Robust} {Deep} {Learning}.
\newblock {\em arXiv:1803.04765 [cs, stat]}, Mar. 2018.
\newblock arXiv: 1803.04765.

\bibitem{poursabzi-sangdeh_manipulating_2019}
F.~Poursabzi-Sangdeh, D.~G. Goldstein, J.~M. Hofman, J.~W. Vaughan, and
  H.~Wallach.
\newblock Manipulating and {Measuring} {Model} {Interpretability}.
\newblock {\em arXiv:1802.07810 [cs]}, Nov. 2019.
\newblock arXiv: 1802.07810.

\bibitem{renkl_toward_2014}
A.~Renkl.
\newblock Toward an {Instructionally} {Oriented} {Theory} of {Example}-{Based}
  {Learning}.
\newblock {\em Cognitive Science}, 38(1):1--37, Jan. 2014. doi: {{%
10\hspace{.1pt}\discretionary{.}{%
}{.}\hspace{.4pt}1111\discretionary{/}{%
}{/}cogs\hspace{.1pt}\discretionary{.}{%
}{.}\hspace{.4pt}12086}}


\bibitem{renkl_example-based_2009}
A.~Renkl, T.~Hilbert, and S.~Schworm.
\newblock Example-{Based} {Learning} in {Heuristic} {Domains}: {A} {Cognitive}
  {Load} {Theory} {Account}.
\newblock {\em Educational Psychology Review}, 21(1):67--78, Mar. 2009. doi:
  {{%
10\hspace{.1pt}\discretionary{.}{%
}{.}\hspace{.4pt}1007\discretionary{/}{%
}{/}s10648\discretionary{%
}{-}{-}008\discretionary{%
}{-}{-}9093\discretionary{%
}{-}{-}4}}


\bibitem{ribeiro_why_2016}
M.~T. Ribeiro, S.~Singh, and C.~Guestrin.
\newblock "{Why} {Should} {I} {Trust} {You}?": {Explaining} the {Predictions}
  of {Any} {Classifier}.
\newblock In {\em Proceedings of the 22nd {ACM} {SIGKDD} {International}
  {Conference} on {Knowledge} {Discovery} and {Data} {Mining}}, pp. 1135--1144.
  ACM, San Francisco California USA, Aug. 2016. doi: {{%
10\hspace{.1pt}\discretionary{.}{%
}{.}\hspace{.4pt}1145\discretionary{/}{%
}{/}2939672\hspace{.1pt}\discretionary{.}{%
}{.}\hspace{.4pt}2939778}}


\bibitem{sannino2018deep}
G.~Sannino and G.~De~Pietro.
\newblock A deep learning approach for ecg-based heartbeat classification for
  arrhythmia detection.
\newblock {\em Future Generation Computer Systems}, 86:446--455, 2018.

\bibitem{selbst}
A.~D. Selbst, D.~Boyd, S.~A. Friedler, S.~Venkatasubramanian, and J.~Vertesi.
\newblock Fairness and abstraction in sociotechnical systems.
\newblock In {\em Proceedings of the Conference on Fairness, Accountability,
  and Transparency}, FAT* '19, p. 59–68. Association for Computing Machinery,
  New York, NY, USA, 2019. doi: {{%
10\hspace{.1pt}\discretionary{.}{%
}{.}\hspace{.4pt}1145\discretionary{/}{%
}{/}3287560\hspace{.1pt}\discretionary{.}{%
}{.}\hspace{.4pt}3287598}}


\bibitem{shin1999memory}
C.~K. Shin and S.~C. Park.
\newblock Memory and neural network based expert system.
\newblock {\em Expert Systems with Applications}, 16(2):145--155, 1999.

\bibitem{sokol2020one}
K.~Sokol and P.~Flach.
\newblock One explanation does not fit all.
\newblock {\em KI-K{\"u}nstliche Intelligenz}, pp. 1--16, 2020.

\bibitem{sturmfels_visualizing_2020}
P.~Sturmfels, S.~Lundberg, and S.-I. Lee.
\newblock Visualizing the {Impact} of {Feature} {Attribution} {Baselines}.
\newblock {\em Distill}, 5(1):10.23915/distill.00022, Jan. 2020. doi: {{%
10\hspace{.1pt}\discretionary{.}{%
}{.}\hspace{.4pt}23915\discretionary{/}{%
}{/}distill\hspace{.1pt}\discretionary{.}{%
}{.}\hspace{.4pt}00022}}


\bibitem{sureshMisplaced}
H.~Suresh, N.~Lao, and I.~Liccardi.
\newblock Misplaced trust: Measuring the interference of machine learning in
  human decision-making.
\newblock In E.~Ferrara, P.~Leonard, and W.~Hall, eds., {\em WebSci '20: 12th
  {ACM} Conference on Web Science, Southampton, UK, July 6-10, 2020}, pp.
  315--324. {ACM}, 2020. doi: {{%
10\hspace{.1pt}\discretionary{.}{%
}{.}\hspace{.4pt}1145\discretionary{/}{%
}{/}3394231\hspace{.1pt}\discretionary{.}{%
}{.}\hspace{.4pt}3397922}}


\bibitem{tison2019automated}
G.~H. Tison, J.~Zhang, F.~N. Delling, and R.~C. Deo.
\newblock Automated and interpretable patient ecg profiles for disease
  detection, tracking, and discovery.
\newblock {\em Circulation: Cardiovascular Quality and Outcomes},
  12(9):e005289, 2019.

\bibitem{tonekaboni2019clinicians}
S.~Tonekaboni, S.~Joshi, M.~D. McCradden, and A.~Goldenberg.
\newblock What clinicians want: Contextualizing explainable machine learning
  for clinical end use.
\newblock In {\em Machine Learning for Healthcare Conference}, pp. 359--380,
  2019.

\bibitem{wachter_counterfactual_2018}
S.~Wachter, B.~Mittelstadt, and C.~Russell.
\newblock Counterfactual {Explanations} without {Opening} the {Black} {Box}:
  {Automated} {Decisions} and the {GDPR}.
\newblock {\em Harvard Journal of Law \& Technology}, 31(2):841--887, Mar.
  2018.
\newblock arXiv: 1711.00399.

\bibitem{wexler2019if}
J.~Wexler, M.~Pushkarna, T.~Bolukbasi, M.~Wattenberg, F.~Vi{\'e}gas, and
  J.~Wilson.
\newblock The what-if tool: Interactive probing of machine learning models.
\newblock {\em IEEE transactions on visualization and computer graphics},
  26(1):56--65, 2019.

\bibitem{zeiler2014visualizing}
M.~D. Zeiler and R.~Fergus.
\newblock Visualizing and understanding convolutional networks.
\newblock In {\em European conference on computer vision}, pp. 818--833.
  Springer, 2014.

\bibitem{zubair2016automated}
M.~Zubair, J.~Kim, and C.~Yoon.
\newblock An automated ecg beat classification system using convolutional
  neural networks.
\newblock In {\em 2016 6th international conference on IT convergence and
  security (ICITCS)}, pp. 1--5. IEEE, 2016.

\end{thebibliography}
\end{document}